\documentclass[article]{jfm}

\usepackage{graphicx}
\usepackage{newtxtext}
\usepackage{newtxmath}
\usepackage{natbib}
\usepackage{hyperref}
\usepackage{soul}
\usepackage{marginnote}

\hypersetup{
    colorlinks = true,   
    urlcolor   = blue,   
    citecolor  = black,
}

\newcommand{\RomanNumeralCaps}[1]
\linenumbers

\usepackage{amssymb,amsmath}
\usepackage{array}
\usepackage{latexsym}
\usepackage{bezier}
\usepackage{psfrag}
\usepackage{bm}
\usepackage{float}
\usepackage{multirow}
\usepackage{mathrsfs}
\usepackage{color}
\usepackage{marginnote}
\usepackage[pdf]{pstricks}
\usepackage{newtxtext}
\usepackage{newtxmath}

\newcommand{\Natural}{\mathbb{N}}

\newcommand{\bvec}[1]{\mbox{\bf #1}}

\newcommand{\bsy}[1]{{\boldsymbol{#1}}}

\graphicspath{{./Figs/}}

\shortauthor{}

\title[Feigenbaum universality in subcritical Taylor-Couette flow]{Feigenbaum universality in subcritical Taylor-Couette flow}

\author[B.\,Wang,  R.\,Ayats, K.\,Deguchi, A.\,Meseguer \and\, F.\,Mellibovsky]{
  B.\,Wang\aff{1},
  R.\,Ayats\aff{1},
  K.\,Deguchi\aff{2} \corresp{\email{kengo.deguchi@monash.edu}},
  A.\,Meseguer\aff{3} \and\,
  F.\,Mellibovsky\aff{3} \corresp{\email{fernando.mellibovsky@upc.edu}}
}

\affiliation{
  \aff{1} Institute of Science and Technology Austria (ISTA), 3400 Klosterneuburg, Austria
  \aff{2} School of Mathematics, Monash University, VIC 3800, Australia 
  \aff{3} Departament de F{\'\i}sica, Universitat Polit\`ecnica de Catalunya, 08034, Barcelona, Spain}

\begin{document}

\maketitle

\begin{abstract}

Feigenbaum universality is shown to occur in subcritical shear flows. Our testing ground is the counter-rotation regime of the Taylor-Couette flow, where numerical calculations are performed within a small periodic domain. The accurate computation of up to the seventh period doubling bifurcation, assisted by a purposely defined Poincar\'e section, has enabled us to reproduce the two Feigenbaum universal constants with unprecedented accuracy in a fluid flow problem. We have further devised a method to predict the bifurcation diagram up to the accumulation point of the cascade based on the detailed inspection of just the first few period doubling bifurcations. Remarkably, the method is applicable beyond the accumulation point, with predictions remaining valid, in a statistical sense, for the chaotic dynamics that follows. 
\end{abstract}

\section{Introduction}\label{sec:introduction}

\begin{figure}
  \begin{center}
    \begin{tabular}{cc}
    (a) & (b) \\
      \includegraphics[height=.4\linewidth]{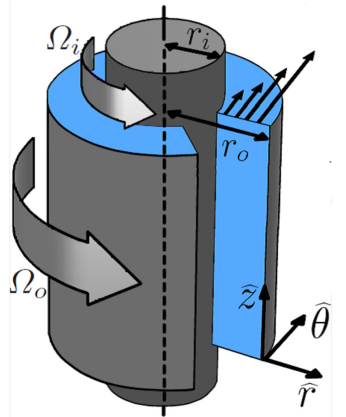} & \includegraphics[height=.4\linewidth]{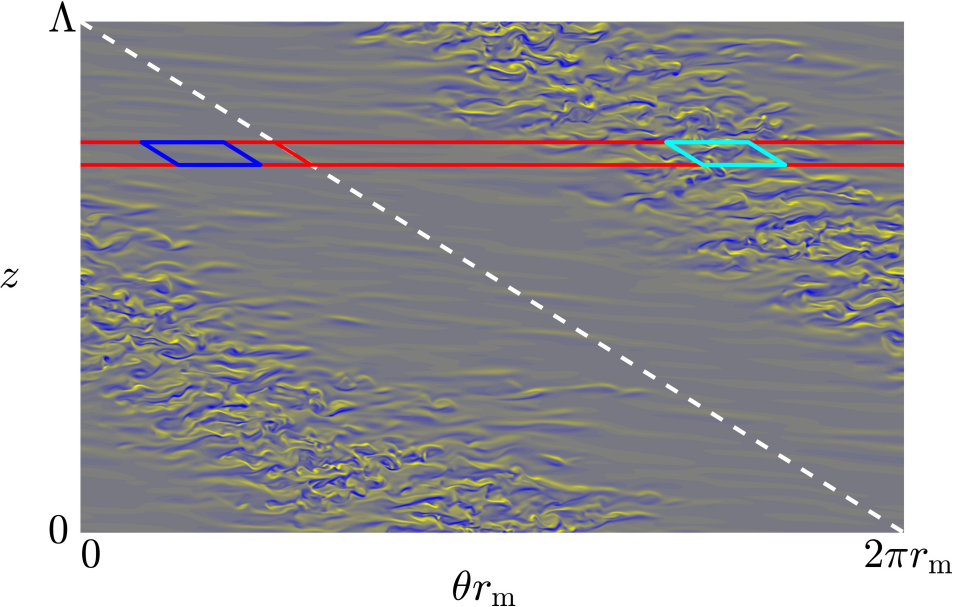} 
      \end{tabular}
 \end{center}  
  \caption{Taylor-Couette flow. (a) Sketch of the flow configuration. The inner and outer cylinders have radii $r=r_i$ and $r=r_o$, respectively, and rotate with angular velocities $\Omega_i$ and $\Omega_o$. (b) A snapshot of the stripe pattern adopted from figure 2 of \cite{WaAyDe22}. The colour map shows the radial vorticity at the mid gap $r_m=(r_o+r_i)/2$. The {radius ratio and inner and outer cylinder Reynolds numbers}, defined in \S\ref{sec:methods}, are set to $(\eta,R_i,R_o)=(0.883,600,-1200)$.}
\label{fig:tcf1}
\end{figure}

The Taylor-Couette flow, driven by two independently rotating coaxial cylinders (figure\,\ref{fig:tcf1}a), has been a prominent research topic in fluid dynamics for over a century. By varying the angular velocities of the cylinders in experimental setups, a rich variety of flow patterns can be observed, as documented by \cite{AnLiSw86}. In this paper, we focus on the case with counter-rotating cylinders, a regime {for which} Taylor-Couette flow has long served as a paradigm {of} subcritical transition to turbulence. The onset of fully developed turbulence is preceded by intermittency, with the flow typically exhibiting alternating laminar and turbulent helical bands \citep{Co65,MeMeAv09a,Do09}; see figure\,\ref{fig:tcf1}b.

The mean flow topology of these helical bands \citep{WaMeAy23} has recently been shown to resemble the laminar-turbulent stripe pattern observed in plane Couette flow \citep{BaTu07}. Indeed, large-scale intermittent patterns of coexisting turbulent and laminar regions are ubiquitous in wall-bounded subcritical shear flows \citep{TuChBa20}. Stochastic approaches, such as directed percolation theory, are gaining popularity {as a framework} for understanding the onset of subcritical turbulent transition \citep{Hof23DP}. However, applying stochastic arguments to a deterministic system inherently assumes the pre-existence of chaotic dynamics. In the case of subcritical shear flows, identifying the emergence of chaos that precedes turbulent transition, poses a significant challenge.

What complicates the detailed understanding of transitional and fully developed turbulence in shear flows, with all the subtle features it encompasses, is the multi-scale nature of the flow. Consequently, many aspects of subcritical parallel shear flow turbulence have historically been approached using small periodic computational domains called \textit{minimal flow units} \citep{JiMo91,HaKiWa95,KaUhlvanVeen12}. 
In Taylor-Couette flow, the use of minimal boxes has been common since the early stages of simulations (e.g. \cite{Marcus84,CoughlinMarcus92}). However, most studies employed parameters in the supercritical regime, and somewhat surprisingly, research in the subcritical regime remains rare. A recent review paper \citep{FeBoBuAvAv23} hypothesised a connection between subcritical Taylor-Couette flow and plane Couette flow, for which \cite{KrEc12} reported a period-doubling cascade leading to chaos in a small periodic box. Here, we confirm that a period-doubling route to chaos does indeed occur in Taylor-Couette flow within the parameter range studied by \cite{MeMeAv09a}, \cite{DeMeMe14}, and \cite{WaAyDe22}, {the latter} hereafter abbreviated as W22.

The most suitable flow unit for the parameter regime we address here is of annular-parallelogram shape proposed by \cite{DeAl13}. W22 first used a narrow elongated domain (red box in figure\,\ref{fig:tcf1}b) {to compute turbulent stripes in their minimal natural box}, a strategy that has been repeatedly adopted in plane Couette \citep{BaTu07} and channel flows \citep{TuKrSc14}. The domain was subsequently shortened in the azimuthal direction (cyan box in figure\,\ref{fig:tcf1}b) so as to fit only one streamwise wavelength of a typical coherent structure of developed turbulence. As reported by W22, the use of such a domain was instrumental in observing the first few period-doubling bifurcations {in the route to chaos}. In this paper, we show that the ensuing period doubling cascade aligns exceptionally well with Feigenbaum universality \citep{Fe78,Fe79,Fe80,Fe82}.

\subsection{Feigenbaum cascade in fluid flows}

Feigenbaum universality became widely recognised among fluid dynamicists following the natural convection experiment conducted by \citet{LiLaFa82}, which led to the first empirical approximation of Feigenbaum's first constant, $\delta_{\rm F} \approx 4.6692016$, corresponding to the limiting ratio of a period-doubling bifurcation interval to the next \citep[see also the recent article by][]{LibchARFM23}.
Their value, $\delta_4=4.4\pm0.1$, estimated from up to the fourth period-doubling bifurcation point (hence the subscript $n=4$), was not too distant from the theoretical value $\delta_{\rm F}$. 
{\cite{BuVoPf93} {found an estimate {close} to $\delta_{\rm F}$ when studying} a Taylor-Couette apparatus with a tilted lid, while subsequent experimental efforts, summarised in table \ref{table_univ}, resulted in {persistently} larger discrepancies.}
\begin{table}
  \begin{center}
    \begin{tabular}{l@{\hspace{.5cm}}l@{\hspace{.5cm}}l@{\hspace{1cm}}l@{ $\pm$ }l@{\hspace{1cm}}l}
      System            & Type & $n$ & \multicolumn{2}{l}{$\delta_n$}  &  Ref. \\\hline
      Thermal convection    & E & $4$ & $4.3 $ & $ 0.8$  &  \cite{GiMuPe81}       \\
      Thermal convection    & E & $4$ & $4.4 $ & $ 0.1$  &  \cite{LiLaFa82}       \\
      Thermosolutal convection   & N-2D & $3$& $5$ & $ ?$  &  \cite{MoToKn83}       \\      
      Taylor-Couette (tilted lid)   & E  & $3$ & $4.67$ & $ ?$    &  \cite{BuVoPf93}  \\
      Baffled channels & E\,\&\,N-2D  & $3$ & $4.65$ & $ ?$    &  \cite{RoMa96}  \\
      Thermal convection & N-2D & $2$ & $4.9104$ & $ ?$ & \cite{LiAle97}  \\
      Triple periodic box & N-3D  & $2$ & $4.88$ & $ 0.5$    &  \cite{Va05}  \\
      Thermal convection & N-3D  & $4$ & $4.321$ & $ ?$    &  \cite{GaPoSe15}  \\
     Forced oscillating cylinder   & N-2D & $5$&$4.52$ & $ ?$  &  \cite{ChJuTo20}       \\
      MHD spherical Couette    & N-3D & $4$&$4.62$ & $ ?$  &  \cite{GaSeGi21}    \\\hline
   \end{tabular}
  \end{center}
  \caption{Feigenbaum universality analyses in fluid systems. The table includes the number of period doubling bifurcations analysed ($n$), and the experimental (E) or numerical (N) nature of the study. An approximation to Feigenbaum's first constant, estimated from the last three period doubling bifurcations analysed in each case, is given in column $\delta_n$.
  }
  \label{table_univ}
\end{table}
Estimates compatible with Feigenbaum's first universal constant have also been obtained numerically for various fluid systems. 
Some of the values reported for $\delta_n$, {also} listed in table\,\ref{table_univ}, are reasonably close to $\delta_{\rm F}$ at first glance.
However, it must be borne in mind that the reliability of the approximations depends strongly on the number of consecutive period-doubling bifurcations analysed and is extremely sensitive to the accuracy with which they are located.
Unfortunately, the approximate determinations of $\delta_{\rm F}$ we have found in the literature of fluid flows were not accompanied with a detailed analysis as done, for example, for the Kuramoto-Sivashinsky equation \citep{SmPa91}. Accordingly, their reliability is debatable, as concluding Feigenbaum universality from a handful of period-doubled solutions is, at the very least, perilous. 

All the studies presented in table\,\ref{table_univ} explore period doubling cascades that follow from a supercritical sequence of bifurcations of the base laminar flow. We focus instead on the {\it subcritical} regime, where turbulent transition may occur despite the linear stability of the base flow and can only be triggered by finite amplitude perturbations. The detection and study of period doubling bifurcations emanating from unstable solutions is impracticable, and stable finite-amplitude solutions in the subcritical regime of shear flows that could potentially originate a cascade are rare and hard to find. 
The aforementioned study by \cite{KrEc12} was the first to show that period-doubling cascades may also occur in subcritical transition problems. Their approach consisted in selecting the smallest possible periodic domain that is capable of sustaining turbulence in plane Couette flow while, at the same time, imposing specific discrete symmetries to further constrain the dynamics.
Unfortunately, the resolution of their parametric exploration was insufficient to positively confirm universality.
Symmetry restrictions were also employed by \citet{MoToKn83} and \citet{Va05} in their respective works. A similar period-doubling cascade in plane Couette flow has also been reported in a more recent paper by \cite{LuKaVa19}. However, calculation of $\delta_{\rm F}$ in the subcritical regime of fluid flow problems has not been conducted to date.

\subsection{Mathematical aspects of the Feigenbaum cascade}

Although period-doubling cascades have been known to occur for over a century \citep{Collet2019}, universality was not discovered until {much later} by \citet{Fe78} and \citet{CouTre78} independently, hence its being often referred to as Feigenbaum-Coullet-Tresser universality {by mathematicians.}  
The existence of universality swiftly spread throughout the mathematical community, as vividly depicted by \citet{KhaLyuSigSin21}. Since its unveiling, universality has been repeatedly observed in numerical simulations of low-dimensional systems of ordinary differential equations. 
Checking universality in low-dimensional models like the R{\"o}ssler system or the Duffing equation has become a common academic exercise. Feigenbaum universality was originally established within the framework of discrete dynamical systems based on one-dimensional iterated maps {of the form }{$x_{\ell+1}=f(x_{\ell})$}{, $\ell \in \mathbb{N}$,} with {$f(x)$} a unimodal function. Its necessary occurrence in continuous time dynamical systems is typically justified by the use of Poincar\'e sections. The phenomenon is often illustrated by initially replacing the Smale horseshoe {that occurs} on the Poincar\'e section with a H\'enon map, which reduces to the logistic map in the limiting case of vanishing area contraction rate after one mapping. Then, since the logistic map is a unimodal map, universality of the period doubling cascade can be explained by means of renormalisation theory, as done in many standard textbooks such as \citet{CoEck80,Schusterbook88,Glenbook94} or \citet{Strogatzbook24}. 
Universality shows in both the parameter (coordinate) and state (ordinate) axes of the bifurcation diagram, the scaling in the latter axis
is related to
Feigenbaum's second constant, $\alpha_{\rm F}\approx -2.5029079$. 
In renormalisation theory, $\alpha_{\rm F}$ and the Feigenbaum function $G$ uniquely solve the 
Feigenbaum-Cvitanovi\'c functional equation $\mathcal{R}[G]=G$ \citep{Fe78,Cvitanovicbook1989}, with the condition $G(0)=1$. 
Here, the action of the renormalisation operator, $\mathcal{R}$, on a function $f$ consists merely in the twice repeated application of $f$, mediated by a rescaling that involves a factor $\alpha_{\rm F}$:
\begin{eqnarray}
 \mathcal{R}[f](x)=\alpha_{\rm F} f \left (f \left (\frac{x}{\alpha_{\rm F}} \right)\right ).\label{eq:RRR}
\end{eqnarray}
Despite its simplicity, this operator lies at the heart of the various {instances of} self-similarity that arise in the bifurcation diagram of the map generated by $f$. Likewise, Feigenbaum's first constant $\delta_{\rm F}$ is the leading eigenvalue of the {linearisation around $G$ of the} renormalisation operator $\mathcal{R}$, and $\varPhi$ is the associated eigenfunction \citep{LanIII82,Ep86,EcWi87,Bri91,StWa91}{, which has recently been dubbed {\it Feigenfunction} by \citet{ThurlbyPhD2021}, as a tribute to Feigenbaum}.
The first mathematical proof of the existence of the universal constants, $\delta_{\rm F}$ and $\alpha_{\rm F}$, and functions, $G$ and $\varPhi$, is attributed to \citet{LanIII82}.

Ascertaining universality from the ability to reproduce $\delta_{\rm F}$ and $\alpha_{\rm F}$ in fluid flow problems presents a twofold challenge.
First, a numerical analysis analogous to that of low-dimensional models demands significantly higher computational resources when dealing with the Navier-Stokes equations. The first observation of $\delta_{\rm F}$ in the field of computational fluid mechanics was contributed by \citet{MoToKn83}, when analysing two-dimensional thermosolutal convection. It took, however, until \cite{Va05} to expose universality in a three dimensional setup, and even then, only for a fluid contained in the simplest computational domain, i.e., periodic in all three space directions.
Second, identifying flow configurations that exhibit the specific route to chaos being targeted, among the several possible in multi-dimensional systems, is much more difficult in physically relevant problems than in engineered toy models with tunable parameters.
 Cascades sequentially doubling the period at every bifurcation may occur through mechanisms quite different from those leading to universality 
 \citep[e.g.,][]{Ya87,KoKoOk96,HoKoNa01,YaWeLo19}. 
Other routes to chaos are also common in fluid systems; 
in the Taylor-Couette system, for example, chaos has been found to arise following the Ruelle-Takens-Newhouse scenario \citep{SwinneyGollub85}
, or Shil'nikov-type bifurcations \citep{LopezPhysD2005}. All these difficulties justify why demonstrating Feigenbaum universality in a fluid system requires careful analysis.

\subsection{The accumulation point of the cascade and beyond}
The orbits of ever increasing period that arise in succession as the parameter is varied along the period-doubling cascade pile up at the {\it accumulation point}, beyond which chaotic dynamics ensue. Our interest extends also to phenomena occurring past this point.
The existence of a reverse cascade, whereby chaotic bands successively merge in pairs as one moves away from the accumulation point, is a well-established  property of simple model maps \citep{GroTho77,Lo80}.
It is noteworthy that \citet{HuRu80} observed self-similarity in the numerical analysis of Lyapunov exponents at the onset of chaos.
Nevertheless, the extent to which universality holds in the chaotic regime remains an elusive question. \citet{Libchaber1983} tried to detect universality beyond the accumulation point in a natural convection experiment. To the authors' knowledge, this is the first and only endeavour to unveil universality past the accumulation point of a period-doubling cascade in a fluid dynamics problem, but the results were rather crude due to experimental limitations. In theory, if universality holds beyond the accumulation point, the bifurcation diagram should be predictable{, not only up to but also past this point,} from the initial few period-doubling bifurcations. However, we have not been able to find any such attempts in the literature, even for low-dimensional models.
While self-similarity of the reverse cascade past the accumulation point has been invariably observed and has been conjectured to hold universally, the underlying mechanisms remain unknown.

\subsection{Outline of the paper}
The paper is structured as follows. The problem formulation is presented in \S\ref{sec:methods}, alongside a brief account of the numerical methods employed. \S\ref{sec:PeriodDoubling} introduces the period doubling cascade that is our object of study and illustrates the procedure by which period doubling bifurcation points are accurately computed. The sequence of the first few such points is then exploited in \S\ref{sec:FeiUni} to assess agreement with the first and second Feigenbaum constants and to extrapolate the expected occurrence of the accumulation point. A method for the detailed prediction of the bifurcation diagram in the neighbourhood of the accumulation point from just a few initial period doubling bifurcations is devised in \S\ref{sec:UniMap}. {We further show how} meaningful predictions can be made, in a statistical sense, that extend beyond the accumulation point into the chaotic regime.
Finally, the main findings are summarised and conclusions drawn in \S\ref{sec:conclusions}.

\section{Formulation of the problem}\label{sec:methods}

We consider an incompressible Newtonian fluid of kinematic viscosity $\nu$, filling the gap between two infinitely long concentric rotating cylinders (figure\,\ref{fig:tcf1}a).
The angular velocities of the inner and outer cylinders, of radii $r^*_{\rm{i}}$ and $r^*_{\rm{o}}$, are denoted as  $\Omega_{\rm{i}}$ and $\Omega_{\rm{o}}$, respectively. A complete set of independent dimensionless physical parameters characterising the problem are the radius ratio $\eta=r^*_{\rm{i}}/r^*_{\rm{o}}$ and the two Reynolds numbers $R_i=dr^*_{\rm{i}}\Omega_{\rm{i}}/\nu$ and $R_o=dr^*_{\rm{o}}\Omega_{\rm{o}}/\nu$, {where $d=r^*_{\rm{o}}-r^*_{\rm{i}}$ is the gap}.

\begin{figure}
  \begin{center}
    \begin{tabular}{cc}
    (a) & (b) \\
\includegraphics[width=.4\linewidth]{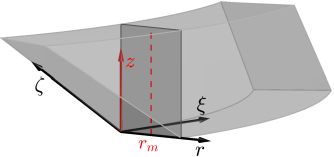} &
      \raisebox{0.87em}{\includegraphics[width=0.4\linewidth]{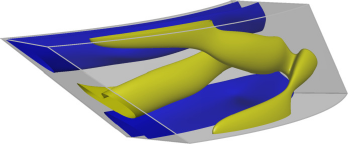}} \\
   \end{tabular} 
 \end{center}  
  \caption{(a) Annular-parallelogram computational domain defined by the coordinates of (\ref{xieta_thetar}) with wave numbers $(n_1,k_1,n_2,k_2)=(10,2,0,4.5)$ and $\eta=0.883$, adopted from W22. The axial line probe (red dashed vertical line) used in the production of space-time diagrams is located at mid gap $r_m=(r_i+r_o)/2 \approx 8.047$. (b) Three-dimensional flow structure of DRW solution for $(R_i,R_o) = (450,-1200)$. Positive (yellow, $u_{\theta} = 250$) and negative (blue, $u_{\theta} = -100$) isosurfaces of perturbation azimuthal velocity.}
\label{fig:drwb}
\end{figure}

All variables are rendered dimensionless using $d$ and $d^2/\nu$ as units for space and time, respectively. In cylindrical coordinates $(r,\theta,z)$, the velocity $\bvec{v}=(v_r,v_{\theta},v_z)=v_r\,\hat{\bsy r} + v_{\theta}\,\hat{\bsy   \theta} + v_z\,\hat{\bsy z}$ and pressure $p$ of the fluid are governed by the Navier-Stokes equations, the incompressibility condition, and the zero axial net massflux condition, i.e.,
\begin{eqnarray}\label{INSE}
  \partial_{t}\bvec{v} + (\bvec{v}\cdot\nabla)\bvec{v} & = & -\nabla p + \nabla^{2}\bvec{v} + f\,\hat{\bsy z},\label{INSE:NS}\\
  \quad \nabla\cdot\bvec{v} & = & 0,\label{INSE:MASS}\\ Q(\bvec{v}) = \int_0^{2\pi}{\int_{r_{\rm{i}}}^{r_{\rm{o}}}{(\bvec{v}\cdot\hat{\bsy z})\,r\,dr\,d\theta}} & = & 0\label{INSE:ZERO},
\end{eqnarray}
where the axial forcing term $f=f(t)$ in (\ref{INSE:NS}) is instantaneously adjusted to fulfil the constraint imposed by (\ref{INSE:ZERO}). The no-slip boundary conditions at the cylinder walls are
\begin{equation}\label{NSBCS}
  \bvec{v}(r_i,\theta,z)=(0,R_i,0),\quad \bvec{v}(r_o,\theta,z)=(0,R_o,0), 
\end{equation}
with $r_i=r_i^*/d=\eta/(1-\eta)$ and $r_o=r_o^*/d=1/(1-\eta)$.
The base, laminar and steady \textit{circular Couette flow}, henceforth referred to as CCF, has
\begin{equation}\label{defccf}
  \bvec{v}_{\rm b} = V_{\rm b} \hat{\bsy \theta} = \displaystyle \left(Ar+\frac{B}{r}\right)\,\hat{\bsy \theta}, \;\; p_{\rm b}(r)=\int\frac{V_{\rm b}^2}{r} \,{\rm d}r, \;\; f_{\rm b} = 0,
\end{equation}
with $ A=(R_o-\eta R_i)/(1+\eta)$ and $B=\eta(R_i-\eta R_o)/\left[(1-\eta)(1-\eta^2)\right]$. In what follows we express the velocity and pressure fields as
\begin{equation}\label{vbpluspert}
  \bvec{v} = \bvec{v}_{\rm b}(r) + \mathbf{u}(r,\theta,z;t), \quad p=p_{\rm b}(r) + q(r,\theta,z;t).
\end{equation}
The fields $q$ and $\mathbf{u}= u_r\,\hat{\bsy r} + u_{\theta}\,\hat{\bsy \theta} + u_z\,\hat{\bsy z}$ are the deviations from the CCF solution.
The nonlinear boundary value problem for $\mathbf{u}$ and $q$ is discretised using a solenoidal Petrov-Galerkin spectral scheme described in \cite{MeAvMe07}. The unknown perturbation fields are approximated by means of a Chebyshev\,$\times$\,Fourier\,$\times$\,Fourier spectral expansion in $(r,\theta,z)$, but in the annular-parallelogram domain $(r,\xi,\zeta) \in \left[r_{\rm{i}},r_{\rm{o}}\right]\times \left[ 0,2\pi \right] \times \left[ 0,2\pi\right]$, where the $2\pi$-periodicity is imposed in the transformed coordinates
\begin{equation}
  \xi  =  n_1\theta + k_1 z,\quad \zeta  = n_2\theta + k_2 z,
\label{xieta_thetar}
\end{equation}
following \citet{DeAl13}. Figure~\ref{fig:drwb}a shows the computational domain corresponding to $\eta=0.883$ and the wavenumber set $(n_1,k_1,n_2,k_2)=(10,2,0,4.5)$.
The domain was specifically designed for the earliest onset of nontrivial solutions while keeping compatibility with the large-scale tilt of spiral turbulence. For this reason, the bifurcation scenario we investigate is fully compatible with the overall structure of intermittent patterns experimentally observed in the counter-rotating regime of the Taylor-Couette system and, at the same time, can be reasonably expected to precede any other mechanism potentially leading to the onset of chaos.
All computations have been run in this domain. The same resolution of $[0,50]\times[-8,8]\times[-8,8]$ modes as used in W22 has been employed throughout. The system of ordinary differential equations {that results from} spatial discretisation {has dimension} $O(10^5)$.

In order to explore the dynamically relevant invariant sets we combine direct numerical simulation (DNS) with a Poincar\'e-Newton-Krylov (PNK) iterative scheme. The spectrally discretised Navier-Stokes equations are integrated in time using a fourth-order linearly implicit IMEX scheme. The PNK scheme, built on top of the time integrator, looks for zeroes of a map defined with a purposely devised Poincar\'e section by means of a Krylov-space-based Newton solver. For a detailed account of the numerical methods used, refer to \cite{AyDeMeMe20} and W22.

We characterise flow states by the \textit{normalised kinetic energy} $\kappa$ of the perturbation velocity field, and by the corresponding inner and outer cylinders \textit{normalised torque}, $\tau_{\rm{i}}$ and $\tau_{\rm{o}}$,
\begin{eqnarray}
   \kappa = \frac{E(\bvec{u})}{E(\bvec{v}_b)},\qquad \tau_{\rm i, \rm o}= \left. 1+\frac{\partial_r(r^{-1}\langle u_{\theta}\rangle_{\xi\zeta})}{\partial_r(r^{-1}V_{b})}\right|_{r=r_{\rm i},r_{\rm o}},
\label{defnormtorque}
\end{eqnarray}
where
\begin{equation}
  E(\bvec{v}) = \frac{1}{2\mathcal{V}} \iiint_\mathcal{V}{\bvec{v}\cdot\bvec{v}\,d\mathcal{V}} =
  \frac{1}{2\mathcal{V}}\int_{0}^{2\pi}\int_{0}^{2\pi}\int_{r_{\rm{i}}}^{r_{\rm{o}}} \bvec{v}\cdot\bvec{v}\,r\,\mathrm{d}r\mathrm{d}\xi\mathrm{d}\zeta =
  \frac{1-\eta}{1+\eta}\int_{r_{\rm{i}}}^{r_{\rm{o}}}{\langle\bvec{v}\cdot\bvec{v}\rangle_{\xi\zeta}\,r\,\mathrm{d}r}
\end{equation}
is the volume-averaged kinetic energy of some velocity field $\bvec{v}$. The volume of the transformed computational domain is $\mathcal{V}=2\pi^2(r_{\rm{o}}^2-r_{\rm{i}}^2)=2\pi^2(1-\eta)/(1+\eta)$, and $\langle~\rangle_{\xi\zeta}$ implies averaging in both parallelogram directions. With these definitions, $\kappa=0$ and $\tau_{\rm{i}}=\tau_{\rm{o}}=1$ for CCF.

The system possesses translational invariance in $\xi$ and $\zeta$. We define the entire set of possible spectral coefficients as the phase space. All coefficient vectors that belong to the group orbit induced by translation of a velocity field represent the same solution. We have systematically factored out the group orbit invariance employing the method of slices \citep{BuCvDaSi15}, using the same template as \cite{WaMeAy23}.
To analyse the period doubling cascade within the framework of discrete-time dynamical systems, we have devised a Poincar\'e section
\begin{equation}\label{eq:PoincSec}
  \Sigma=\left\{{\bf a}\in \mathbb{X}\left |\;\tau_i({\bf a})=\tau_o({\bf a}),\;\dfrac{d\tau_i}{dt}>\dfrac{d\tau_o}{dt}\right. \right\},
\end{equation}
in phase space $\mathbb{X}$, which consists of all possible sliced spectral expansion coefficients vector $\mathbf{a}$.
The condition defining $\Sigma$ is based on the equality of inner and outer cylinder torque (first condition) and the sign of the rate of change of their difference (second condition) to discard reverse crossings. 
It is easy to check from the governing equations that for statistically steady states the time average of $\tau_{i}$ must coincide with that of $\tau_{o}$. {Consequently,} long-lived (including permanent) time-dependent solutions are bound to regularly fulfil the condition $\tau_{\rm{i}}=\tau_{\rm{o}}$, a property that comes in handy in defining a robust Poincar\'e section.
To simplify notation, we will hereafter call $\tau=\tau_i=\tau_o$ the value of the torque, whether inner or outer, on $\Sigma$.

There are other ways of sampling a continuous system to produce a discrete system. 
For example. \cite{KrEc12} and \cite{GaPoSe15} investigated the relationship between consecutive local maxima of the time series of some physical quantity.
However, it is not easy to discern which minima/maxima are actually related to the underlying (pseudo-)periodicity of the solutions and which are simply incidental to the dynamics of the particular time signal used. 
Thus the {use} of the above tailor-made Poincar\'e section provides a more reliable approach.

\section{The period doubling cascade}\label{sec:PeriodDoubling}

Emulating the influential experiments by \cite{AnLiSw86}, we employ a radius ratio $\eta=0.883$ ($r \in [r_i,r_o]=[7.547,8.547]$). The outer cylinder Reynolds number is fixed at $R_o = -1200$. 
For sufficiently large $R_i$, the centrifugal instability of the base flow develops into full-fledged turbulence. A reduction of $R_i$ while still in the supercritical region has the flow re-organise into a pattern of alternating laminar and turbulent winding helical bands, commonly known as the spiral turbulence regime \citep{Co65,AnLiSw86,MeMeAv09a,Do09}. At this particular value of $R_o$, the CCF is linearly stable for $R_i \lesssim 447.35$,  
but a number of subcritical nonlinear equilibrium states persists below this threshold \citep[][and W22]{MeMeAv09b,DeMeMe14}, which are believed to act as precursors of spiral turbulence.
Drawing from the helix angle of the turbulent stripes observed experimentally, W22 employed the domain and coordinate system shown in figure \ref{fig:drwb}a to seek minimal flow unit solutions that are compatible with the tilt of the banded pattern. Subharmonic instabilities of some such solutions were proposed as the mechanism that might be responsible for the spatial intermittency characterising spiral turbulence. Within this minimal parallelogram-annular periodic domain, which is capable of sustaining turbulence at sufficiently high $R_i$, W22 identified {a stable finite amplitude travelling wave (shown in figure\,\ref{fig:drwb}b) that coexists with the stable CCF. This drifting rotating wave, DRW for short, embodies} all the essential elements of the self-sustaining mechanism \citep{Wang2007,HallSherwin2010} and plays a central role in the transition process. In W22, two periodic solutions, P$_1$, arising from a Hopf bifurcation of DRW, and P$_2$, originating at a period-doubling bifurcation of P$_1$, were also identified as the first steps in the route to chaos.
This sequence of bifurcations suggested a period-doubling cascade as the most plausible scenario for the onset of chaotic dynamics, but the issue was not pursued further.

\subsection{The onset of the period doubling cascade}\label{sec:start}

We now look into the bifurcation sequence the system undergoes in the range $R_i\in[395.43,395.79]$ eventually leading to sustained chaotic dynamics. Since we will only be varying $R_i$, we shall drop the subscript and simply write $R$ to ease notation.

\begin{figure}
\begin{center}
\raisebox{0.33\linewidth}{(a)} \includegraphics[height=.365\linewidth]{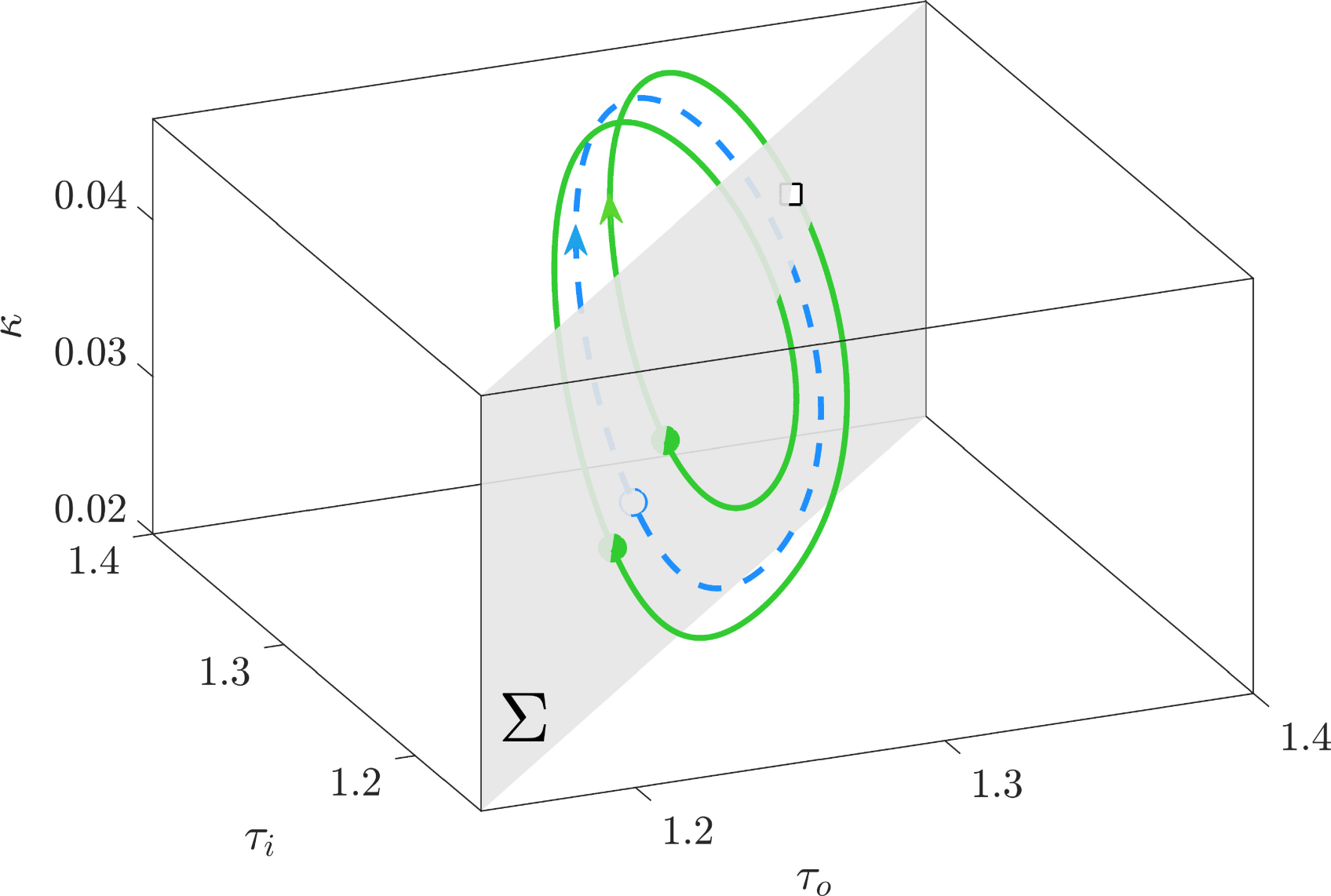}
  \begin{tabular}{l}
    \hspace{0.10\linewidth} (b) \hspace{0.61\linewidth} (c) \\
    \includegraphics[width=.95\linewidth]{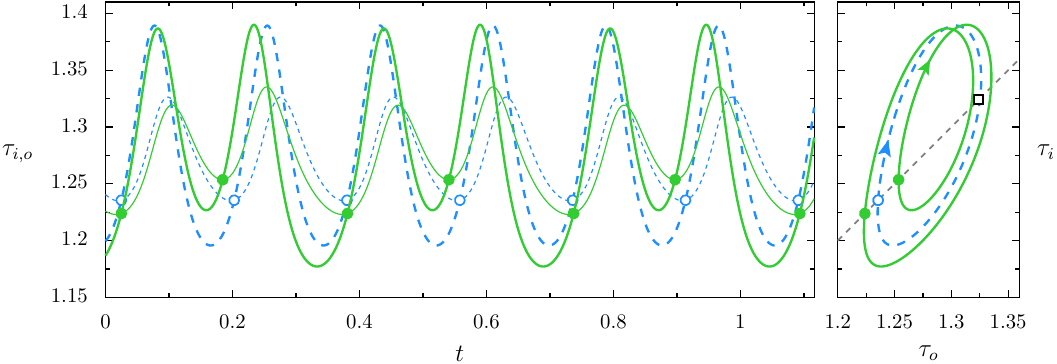}
   \end{tabular} 
 \end{center}  
  \caption{The Poincar\'e section $\Sigma$ and the periodic orbits P$_1$ (dashed blue) and P$_2$ (solid green). The black squares are DRW solutions. All solutions are computed at $R=395.67$. (a) Projection of the phase space on the $(\tau_o,\tau_i,\kappa)$ coordinates. (b) Inner ($\tau_i$, thick lines) and outer ($\tau_o$, thin) torque time series of P$_1$ and P$_2$. (c) Two-dimensional phase map projection on the $(\tau_o,\tau_i)$ plane. The Poincar\'e section is shown in transparent grey in panel (a) and as a dashed grey line in panel (c). The circles on the P$_1$ (empty blue) and P$_2$ (filled green) curves correspond to their representation on $\Sigma$.
  }\label{fig:P1P2}          
\end{figure}

Figure~\ref{fig:P1P2}a shows the DRW (square), P$_1$ (dashed blue line) and P$_2$ (solid green line) at $R=395.67$ in a three-dimensional projection of the full phase space  $\mathbb{X}$ on the subspace spanned by the triplet of key quantities $(\tau_o,\tau_i,\kappa)$.
The Poincar\'e section $\Sigma$ appears in this representation as the transparent grey plane, which contains DRW and is pierced, in the direction defined by \eqref{eq:PoincSec}, at a single point by P$_1$ and at two different points by P$_2$.
In the torque time series of figure~\ref{fig:P1P2}b, the Poincar\'e crossings are conveniently identified as the intersections between the $\tau_i$ (thick lines) and $\tau_o$ (thin) signals for which the former is overtaking the latter. The two-dimensional projection of figure~\ref{fig:P1P2}a onto the plane $(\tau_o,\tau_i)$, as depicted in figure~\ref{fig:P1P2}c, further clarifies how the sequence of intersection points of figure~\ref{fig:P1P2}b collapses in phase space onto a single point for P$_1$ (empty blue disk) and as two distinct points for P$_2$ (filled green circles), all contained in $\Sigma$ (dashed grey straight line).

\begin{figure}                                                                 
  \begin{center}
    \begin{tabular}{l@{}cl@{}c}
      \raisebox{0.06\columnwidth}{(ai)} & \includegraphics[height=.082\linewidth]{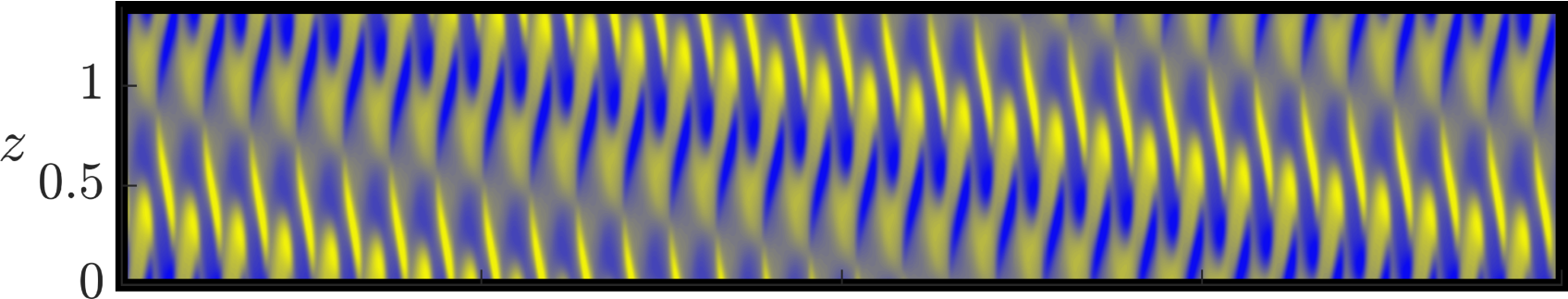} 
      &
      \raisebox{0.06\columnwidth}{(bi)} & \raisebox{0.002\columnwidth}{\includegraphics[height=.0805\linewidth]{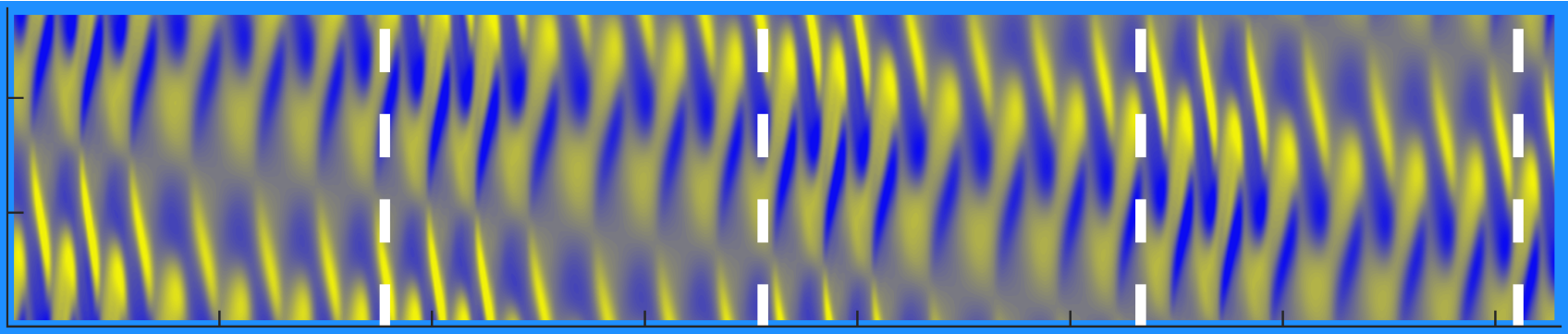}}
      \\[-0em]
      \raisebox{0.06\columnwidth}{(aii)} & \includegraphics[height=.082\linewidth]{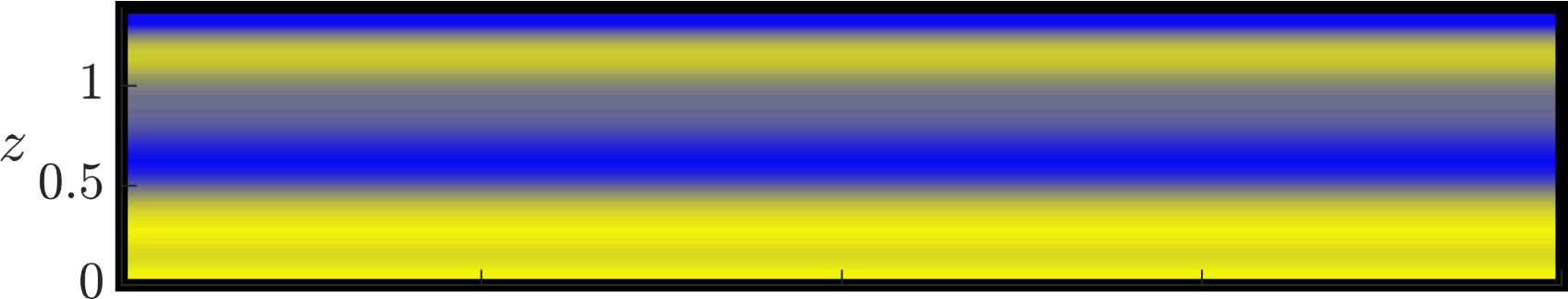} 
      &
      \raisebox{0.06\columnwidth}{(bii)} & \raisebox{0.002\columnwidth}{\includegraphics[height=.0805\linewidth]{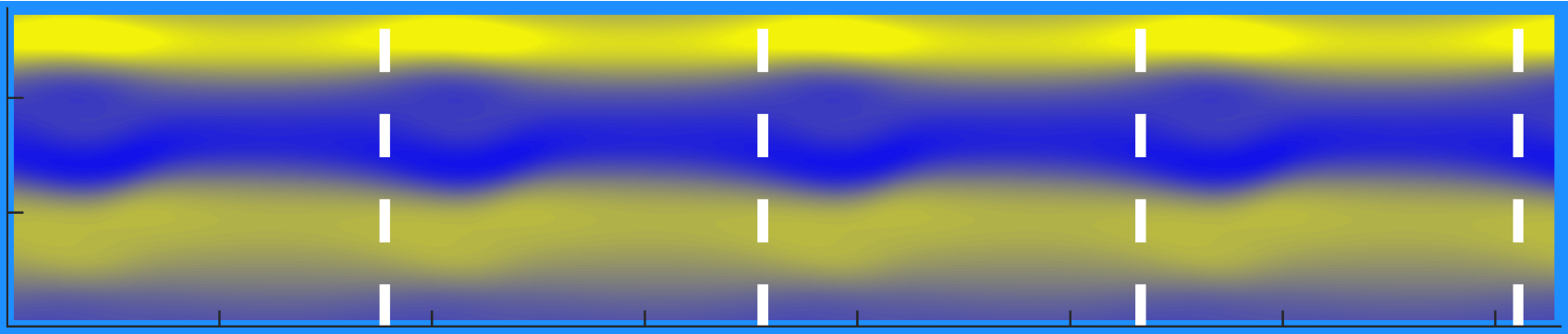}}
      \\[-0em]
      \raisebox{0.09\columnwidth}{(ciii)} & \includegraphics[height=.115\linewidth]{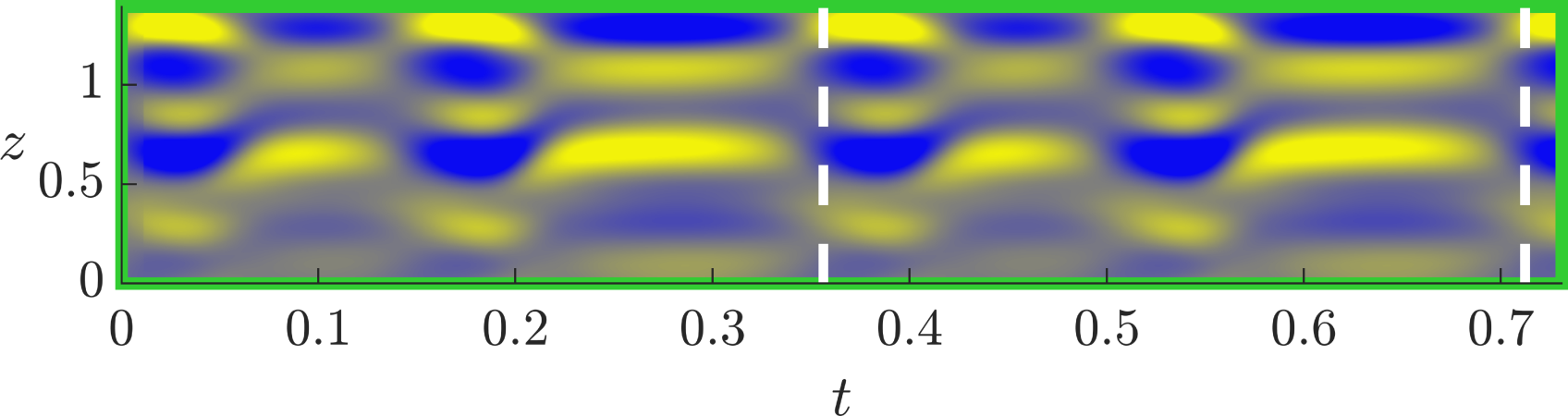} 
      &
      \raisebox{0.09\columnwidth}{(biii)} & \includegraphics[height=.109\linewidth]{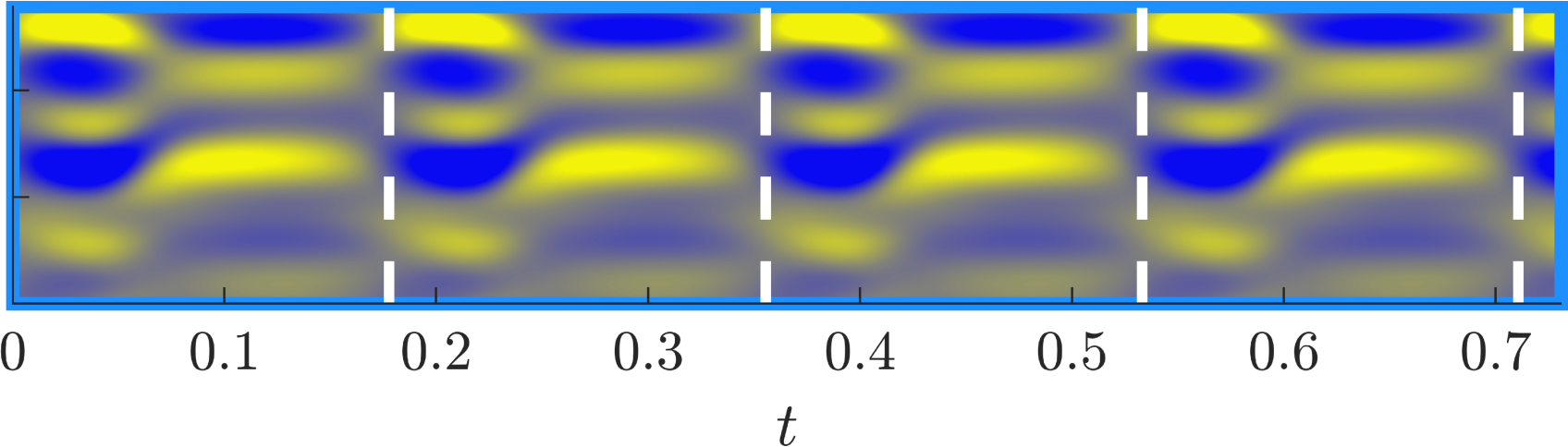} 
      \\
    \end{tabular} 
  \end{center}  
  \caption{Space-time diagrams for (a) DRW, (b) P$_1$ and (c) P$_2$, all computed at $R=395.67$. The roman number labels denote measurements of radial vorticity $\omega_r(z;t)$ along axial probe lines at $(r,\theta)=(r_m,\theta_0)$ fixed to (i) the lab (stationary) reference frame, (ii) a reference frame co-moving with the solution, and (iii) the same co-moving frame but with the temporal mean $\langle\omega_r\rangle_{t}$ subtracted. The azimuthal location, $\theta_0$, is chosen consistently across reference frames and solutions to enable comparison. Colour shading according to $\omega_r\in[-1400,1400]$ or $\omega_r- \langle\omega_r\rangle_{t}\in[-300,300]$, as need be. Dashed vertical lines indicate the natural period of the corresponding solution.
}
\label{fig:spacetime}          
\end{figure}

The drift dynamics of non-axisymmetric solutions in Taylor-Couette as we have here is inevitably masked when monitoring aggregate quantities such as torque or kinetic energy due to their spatial averaging properties. Point measurements are instead subject to drift-induced time dependence. Figure \ref{fig:spacetime}ai presents a space-time diagram of the axial vorticity distribution measured along a line probe that is fixed in the lab reference frame (red dashed line in figure \ref{fig:drwb}a).
DRW, a relative equilibrium, is characterised by a solid-body motion composing axial translation and azimuthal rotation with constant phase velocities. As a result, the space-time diagram exhibits the repetition of a periodic pattern.
The same line probe produces a quasi-periodic space-time diagram for P$_1$, as shown in figure~\ref{fig:spacetime}bi. P$_1$ is a relative periodic orbit and, therefore, requires suitable shifts in both the $z$ and $\theta$ directions to align the flow fields after every one period to reveal the space-time invariance. The effects of the drift can be suppressed by attaching the line probe to a moving reference frame defined with the method of slices \citep{BuCvDaSi15,WaMeAy23}. In this reference frame, the space-time diagrams of DRW and P$_1$ (figures \ref{fig:spacetime}aii and \ref{fig:spacetime}bii) are greatly simplified, the former appearing as time-independent and the latter as purely time-periodic.

Further subtraction of the time average helps expose the true nature of the time dependence of the solutions. For P$_1$, the time-horizon considered in figure~\ref{fig:spacetime}biii allows for just over four repetitions, as indicated by the dashed vertical lines. Solution P$_2$, shown in figure~\ref{fig:spacetime}ciii, has instead a natural period about twice that of P$_1$, the period-doubling consisting in a modulation that shortens one of the {\it half}-periods while lengthening the other.
  
\begin{figure}                                                                 
  \begin{center}
    \begin{tabular}{cc}
      (a) & (b) \\
      \includegraphics[height=.365\linewidth]{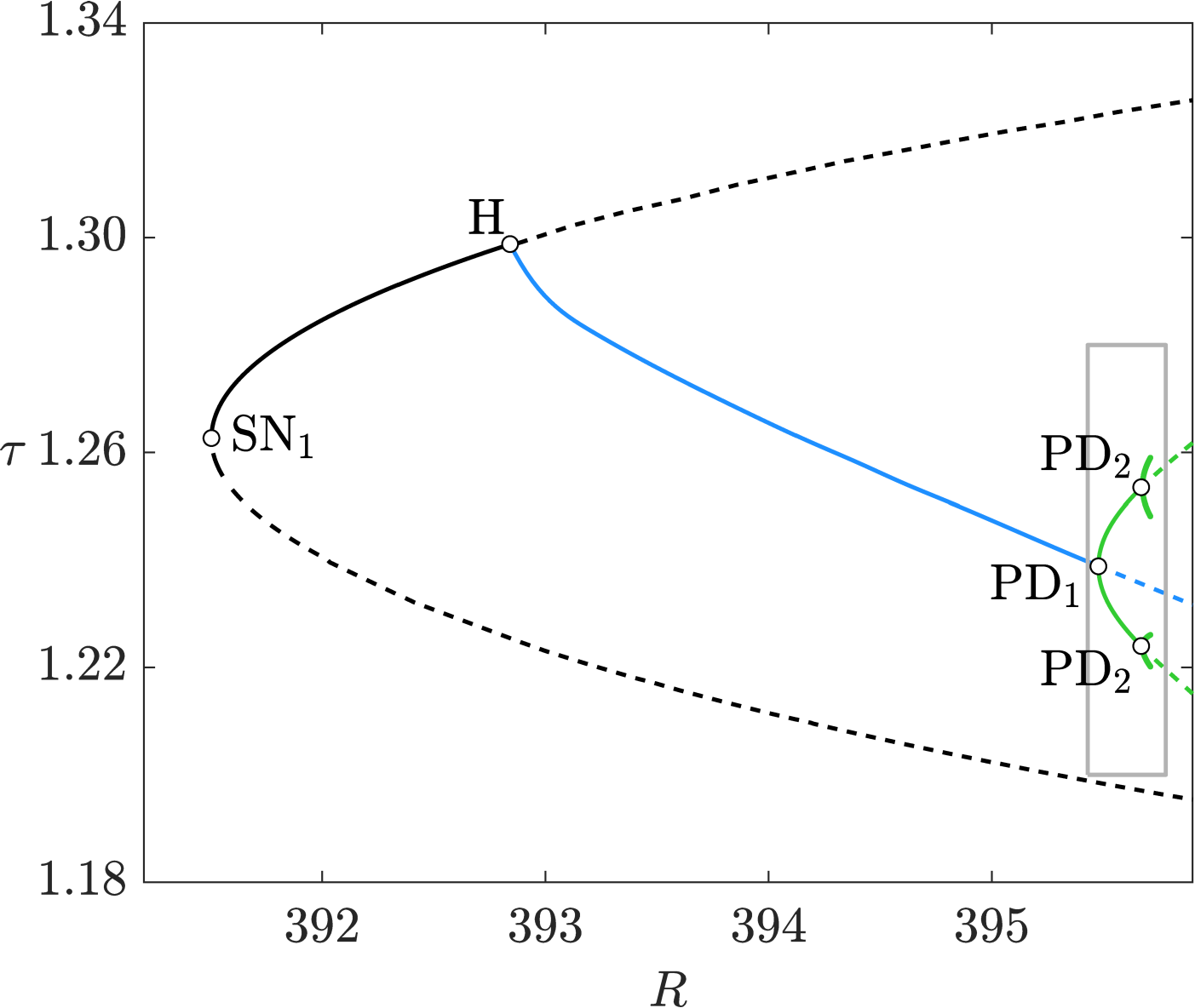} &
      \raisebox{.4em}{\includegraphics[height=.36\linewidth]{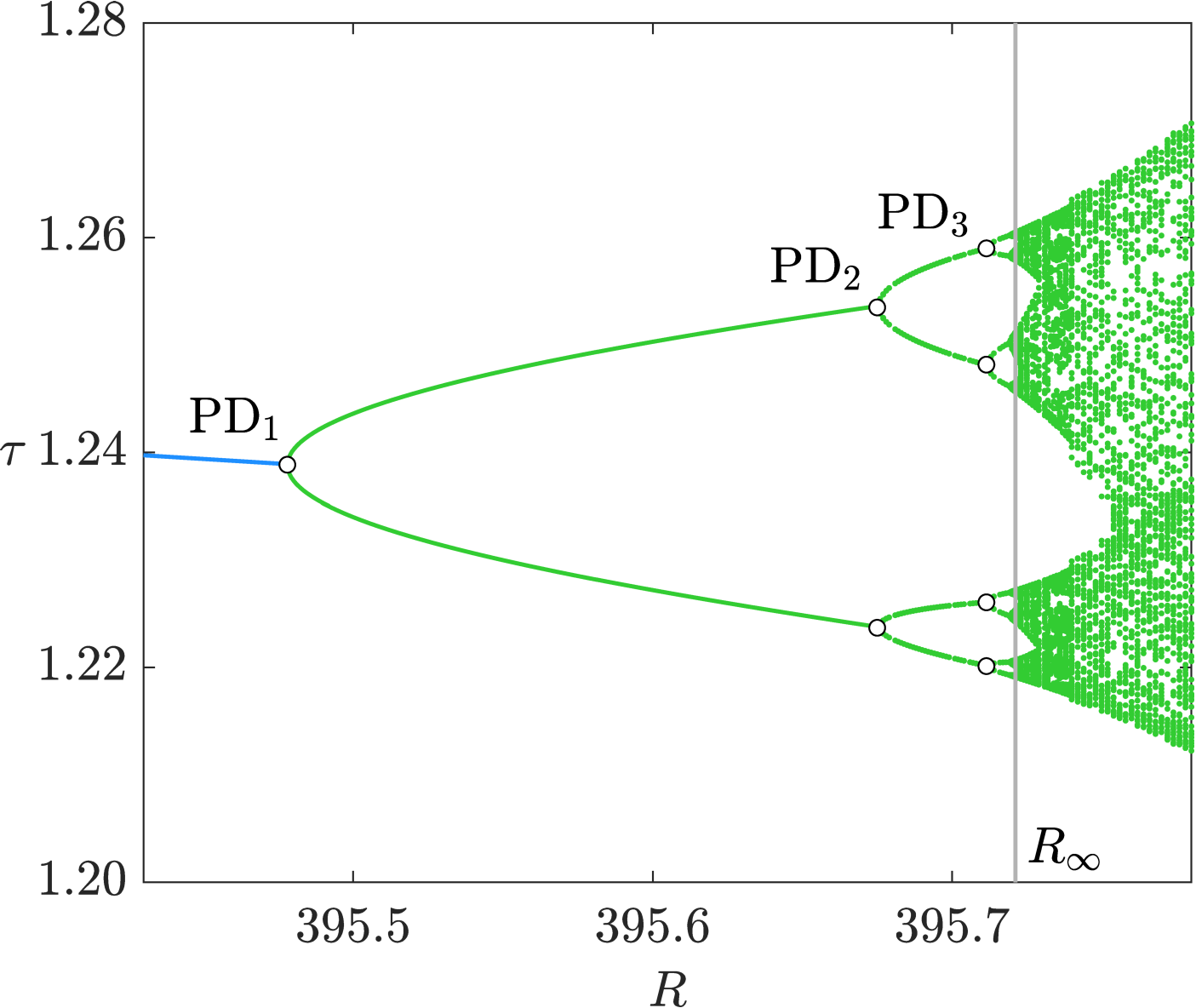}} \\
    \end{tabular} 
  \end{center}  
  \caption{Bifurcation scenario as recorded on the Poincar\'e section $\Sigma$. (a) The initial steps of the bifurcation scenario. Shown are DRW (black), P$_1$ (blue) and P$_2$ (green), reported in W22. P$_4$ (green) emerges at the second period doubling bifurcation point PD$_2$. Both stable (solid line) and unstable (dashed) solution branches are shown. (b) Detailed view (close-up of the region bounded by a solid grey box in panel a) of stable solution branches across the period-doubling cascade and beyond. The accumulation point for the period doubling cascade ($R_\infty$) is to be computed in \S\ref{sec:FeiUni}.}
  \label{fig:biffigP1}          
\end{figure}

Let us now shift our focus to the dependence of the solutions on the parameter $R$. Figure~\ref{fig:biffigP1}a shows the bifurcation sequence that DRW undergoes when varying the inner Reynolds number in the range {$R\in[391.20,395.90]$}. 
DRW emerges from a saddle-node bifurcation (SN$_1$) at $R\simeq391.50$ and, leaving aside subharmonic instabilities, the nodal (upper) branch (the one represented in figure~\ref{fig:drwb}b and the square in figures~\ref{fig:P1P2}a,c) remains stable to perturbations fitting the domain until the advent of a supercritical Hopf bifurcation (H) at $R\simeq392.85$. As reported by W22, this is the only known case of a {
stable
non-trivial solution in the subcritical parameter regime of Taylor-Couette flow.
Note that here we are only interested in the stability to perturbations that fit within the periodic box. Stable/unstable solutions are denoted by solid/dashed lines. The PNK method must be used to compute unstable solution branches.
A stable relative periodic orbit (P$_1$, blue line) bifurcates from DRW at H. All time-dependent solutions will be represented, from now on, through the collection of intersection points on the Poincar\'e section. The P$_1$ branch becomes unstable in a period doubling bifurcation (PD$_1$), whence a branch of stable period-doubled relative periodic orbits (P$_2$, pair of green lines) emerges. The P$_2$ loses stability, in turn, in a second period-doubling bifurcation (PD$_2$) issuing a branch of period-4 solutions. The $R$ value for which figures~\ref{fig:P1P2} and \ref{fig:spacetime} were computed corresponds to just short of PD$_2$, hence the instability of DRW and P$_1$, in contrast with P$_2$, which is stable.

The remainder of the period-doubling cascade and the onset of chaotic dynamics at larger values of $R$ is depicted in figure~\ref{fig:biffigP1}b. Only stable solutions are shown, so DNS is all that has been required to produce the solution branches in the diagram. Checking agreement with $\delta_F$ from the approximate location on the bifurcation diagram of the period doubling points may seem a straightforward task. However, confirmation of Feigenbaum universality by brute force poses challenges, even for simple systems such as the logistic map. To begin with, the parameter spacing between consecutive period-doubling bifurcations shrinks very fast as one progresses along the cascade, demanding the computation of bifurcation points with an ever increasing number of significant digits. {Moreover,} the orbits in the immediate vicinity of a period doubling point, which are required for the accurate estimation of the bifurcations, are close to neutrally stable and, therefore, take massive computational time to reach convergence with DNS. Unfortunately, employing the PNK method becomes impractical for solutions of increasingly long periods.
 The inability of the method to discriminate between stable and unstable orbits, combined with the fact that stable orbits in a period doubling cascade coexist with all unstable orbits of lower period, which are favoured unless very close initial guesses are produced in advance, curbs any advantage one might have expected to achieve from using PNK.
Both complications combined render the careful appraisal of agreement with Feigenbaum's first constant extremely hard. A rigorous and systematic analysis as detailed in the coming section becomes thus necessary.

\subsection{Determination of period doubling points}\label{sec:determination}

To analyse the period-doubling cascade, it is most convenient to focus on the sequence of ${\bf a}$ on $\Sigma$. In the framework of dynamical systems theory, this corresponds to investigating the properties of the Poincar\'e map{, also known as first return map}. As we shall explain later, it is not necessary to monitor the full set of coefficients, and the sequence of torque values provides (nearly) all necessary information in the long run, once the initial transients are over.
Let $\tau(\ell)$, $\ell=1,2,3,\dots$ be the corresponding sequence of torque values on $\Sigma$, with $\ell$ an index recording the chronological order.
As an example, figure~\ref{fig:P4P8}a shows $\tau(\ell)$ for the P$_4$ solution (red circles) at $R=395.711$ (between PD$_2$ and PD$_3$) and for the P$_8$ solution (blue) at $R=395.719$ (between PD$_3$ and PD$_4$).
\begin{figure}                                                                 
  \begin{center}
  \begin{tabular}{l}
    \hspace{0.10\linewidth} (a) \hspace{0.53\linewidth} (b) \\
    \includegraphics[width=.95\linewidth]{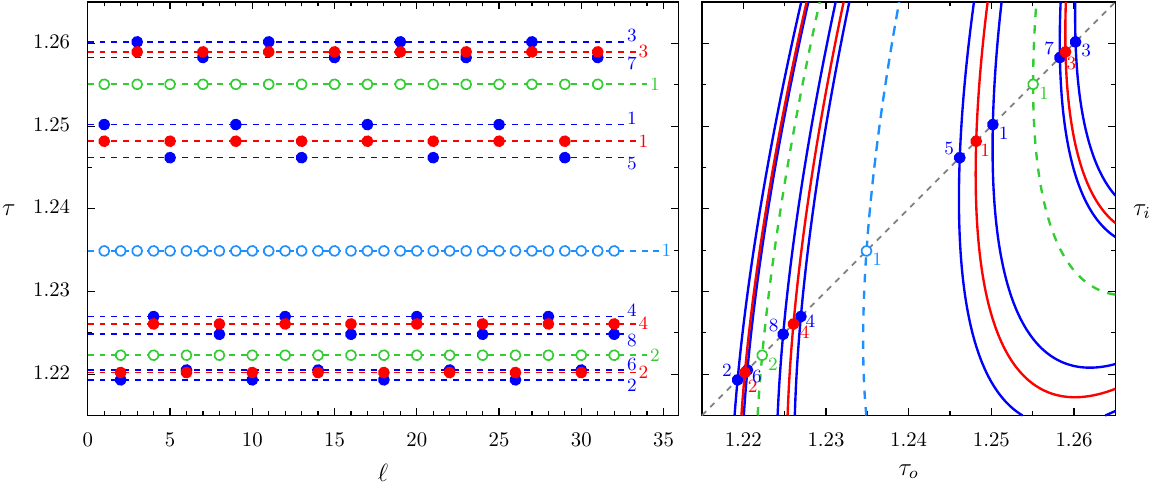}
   \end{tabular} 
 \end{center}  
  \caption{All orbits up to period 8 at (P$_1$ and P$_2$, unstable) and around (P$_4$ and P$_8$, at $R=395.711$ and $395.719$, respectively) PD$_3$.   
(a) Torque ($\tau=\tau_i=\tau_o$) of P$_1$ (cyan), P$_2$ (green), P$_4$ (blue) and P$_8$ (red) on $\Sigma$ as a function of the discrete time $\ell$ (crossing index). The dashed lines indicate the distinct values of $\tau$. (b) Two-dimensional phase map projection on the $(\tau_o,\tau_i)$ plane. Shown are the phase map trajectories of all four orbits (dashed line for unstable, solid for stable) along with their representation on the Poincar\'e section (circles, open for unstable, filled for stable). The numbers indicate the order of the crossings.
  }\label{fig:P4P8}          
\end{figure} 
Both solutions are stable and well converged with DNS, such that the respective series sequentially repeat, always in the same order, the same four (P$_4$) and eight (P$_8$) distinct values, as indicated by the numbered horizontal dashed lines.
The effect of the period-doubling bifurcation PD$_3$ can thus be portrayed as doubling point $j$ of the P$_4$ discrete-time orbit into points $j$ and $j+4$ of the P$_8$ cycle, the latter two points having sprung from the former and drifted away in opposite directions. 
The unstable P$_1$ (cyan) and P$_2$ (green) orbits {in the immediate vicinity of} PD$_3$ are shown for reference, to help generalise the rule that relates points $j$ and $j+N/2$ of the period-$N$ orbit to point $j$ of the period-$N/2$ orbit from which they bifurcate. A phase map analogous to that of figure~\ref{fig:P1P2}c is shown in figure~\ref{fig:P4P8}b, but with the region where $\Sigma$ is pierced by the orbits suitably
magnified to further clarify where the P$_4$ and P$_8$ cycles stand in relation to P$_1$ and P$_2$.

The $n$th period doubling bifurcation PD$_n$ can be readily analysed by splitting the $\tau(\ell)$ sequences of interest into $J=2^n$ separate subsequences
\begin{equation}\label{eq:kNj}
\tau_j^J(k)=\tau(kJ+j),
\end{equation}
each starting at one of $J$ consecutive Poincar\'e crossings $j\in\{1,\ldots J\}$, and sampling every $J$th crossing thereafter.
The limits
\begin{equation}\label{eq:tauinfty}
  (\tau_j^J)_\infty = \lim_{k\to\infty}\tau_j^J(k) \qquad j\in\{1,\ldots J\}
\end{equation}
exist for all subsequences of all orbits of period $N=J=2^n$ or lower ($J/2,J/4,\dots$) along the cascade, but fail for orbits of longer period ($2J,4J,\dots$). Moreover, only period-$J$ orbits will produce $J$ distinct $(\tau_j^J)_\infty$ values, while shorter period orbits will have the limits coincide in pairs according to $(\tau_j^J)_\infty=(\tau_{j+J/2}^J)_\infty$. This property is instrumental in telling apart orbits of period $N=J$ from orbits of period $J/2$ or smaller, as the amplitudes defined by
\begin{equation}\label{eq:amplitude}
  A_j^J = \left\lvert(\tau_j^J)_\infty-(\tau_{j+J/2}^J)_\infty\right\rvert \qquad j\in\{1\ldots J/2\},
\end{equation}
will all vanish for any orbit other than P$_J$. For instance, suppose we slightly perturb the stable P$_8$ at $R=395.719$. If we sample the $\tau$ sequence with $J=8$, all subsequences $j=1,2,3,\dots,8$ converge and all eight limits are different (recall figure~\ref{fig:P4P8}). Applying the same $J=8$ sampling to a stable P$_4$ (or P$_2$ or P$_1$) will instead produce indistinguishable limits for $j$ and $j+4$ and, consequently, vanishing amplitudes $A_j^8=0$.

To nail down PD$_3$, a collection of DNS runs traversing the bifurcation point, i.e., between $R=395.711$ and $395.719$, are required. The amplitude of P$_8$ drops fast as the parameter $R$ is decreased towards PD$_3$, such that  distinguishing it from P$_4$ becomes increasingly difficult. In addition, the dynamics become despairingly slow as the bifurcation point is approached from either side, rendering convergence with DNS downright impracticable. Figures~\ref{fig:PowFitPD8}a and \ref{fig:PowFitPD8}b depict the torque subsequences at $R=395.7116$ and $R=395.7122$, respectively, below and above PD$_3$ but {moderately} close to the bifurcation point.
\begin{figure}
  \centering
  \begin{tabular}{c@{\hspace{0.5em}}c}
    (a) & (b) \\
    \includegraphics[height=.75\linewidth]{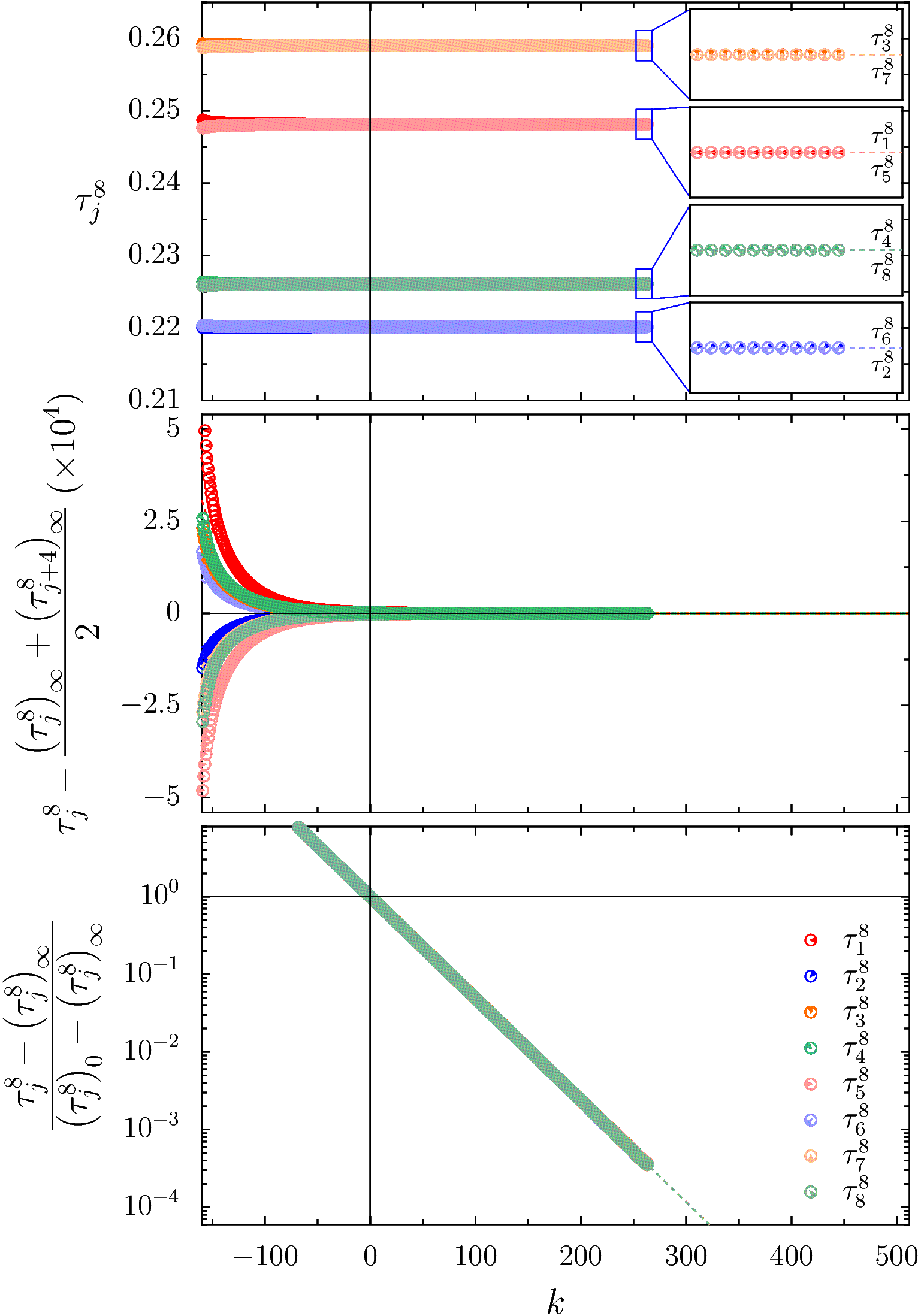} &
    \includegraphics[height=.75\linewidth]{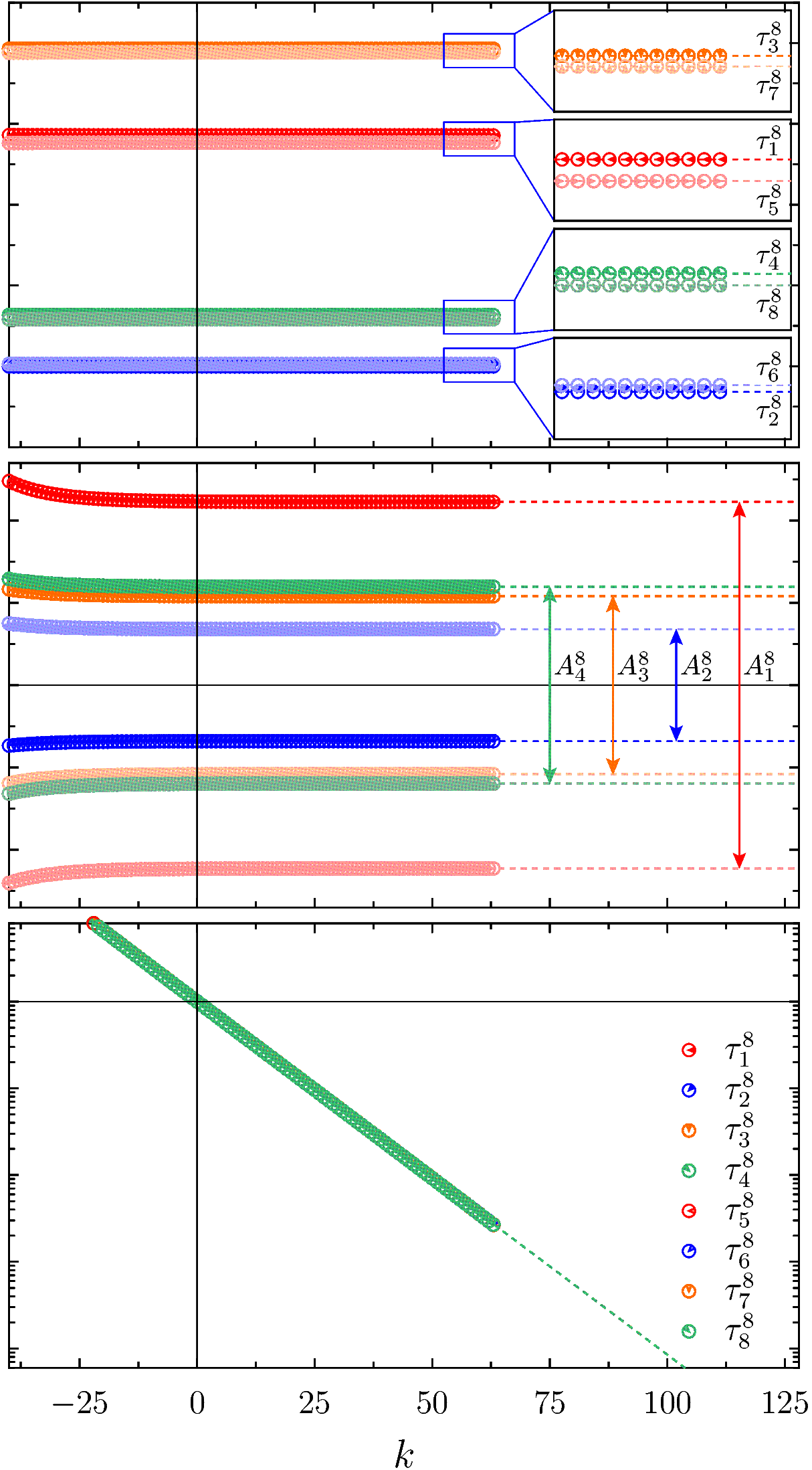}
  \end{tabular}
  \caption{Analysis of torque sequences $\tau(\ell)$, sampled with $N=8$ at (a) $R=395.7116$ and (b) $R=395.7122$, below and above PD$_3$ ($R=395.711841564$, as we shall see later), respectively. The top panels show the 8 sequences $\tau_j^8(k)$ (circles, magnified in the insets), alongside respective power law fits (dashed lines). To differentiate the 8 subsequences, the points corresponding to each are coloured and represented with disks that are mainly blank except for a 1/8th sector, whose orientation ($360^\circ \times j/8$) uniquely identifies the corresponding sequence $\tau_j^8$ (see the legend). The mid panels show the same data but with the mean value $((\tau_j^8(k))_\infty+(\tau_{j+4}^8(k))_\infty)/2$ subtracted from every pair of branches to illustrate convergence. The bottom panels show again the same data but in logarithmic scale.
}\label{fig:PowFitPD8}
\end{figure}
Every pair of sequences ($\tau_j^8$ and $\tau_{j+4}^8$, 
depicted in different shades of the same colour) appears to be converging to the same value at $R=395.7116$ but not at $R=395.7122$. At these values of $R$ the achievable degree of convergence is still reasonable but sufficiently slow to illustrate how the estimation of the amplitudes may be systematised to parameter values much closer to the bifurcation point.
{Well past initial transients, and as the sequences approach convergence, the dynamics is progressively determined by the governing equations linearised around the stable solution. Therefore, we can expect that a power law fit of the form 
\begin{equation}\label{eq:PowFit}
  \tau_j^J(k) =(\tau_j^J)_{\infty}+ \left((\tau_j^J)_0-(\tau_j^J)_{\infty}\right)(\lambda_j^J)^k,
\end{equation}
with fitting parameters $(\tau_j^J)_{\infty}, (\tau_j^J)_0,$ and $\lambda_j^J$, {conforms reasonably well} to the final transients. The fit provides an estimate for both the asymptotic torque value $(\tau_j^J)_{\infty}$ and the dominant multiplier $\lambda_j^J$.}
{The dashed lines in the insets of the top panels of figure~\ref{fig:PowFitPD8} are the fits to the sequences $\tau_j^J(k) $, $k\geq 0$.} 
In order to expose the convergence rate and the accuracy of the fits, the same data has been plotted in the mid panels, now subtracting from each pair of branches $j$ and $j+4$ the mean value $((\tau_j^8)_\infty+(\tau_{j+4}^8)_\infty)/2$ at infinity. The fits provide an excellent approximation to the data points because the dynamics have already reached the linear regime. While all fitting curves decay to zero for $R=395.7116$ in this representation, four distinct amplitudes arise for $R=395.7122$. This unequivocally identifies the former dynamics as converging on a P$_4$ and the latter on a P$_8$. The bottom panels plot the same data yet again, but now in logarithmic scale. In this representation, all fits to the individual torque subsequences are seen to collapse onto a single straight line.
The eight slopes $\lambda_j^8$, indistinguishable from one another to the naked eye, provide an estimate to the dominant multiplier of the converging periodic solution, interpreted as a P$_8$ regardless of its actual period.

The four amplitudes $A_j^8$ and eight multipliers $\lambda_j^8$ have been computed, following the same procedure with the sampling $J=8$, at several values of $R$ approaching PD$_3$ from either side, and plotted in figure~\ref{fig:PD8}.
\begin{figure}
  \centering
  \begin{tabular}{l@{}c}
    \raisebox{0.42\linewidth}{\parbox{2em}{(a)\\[13em](b)}} &
    \includegraphics[height=.6\linewidth]{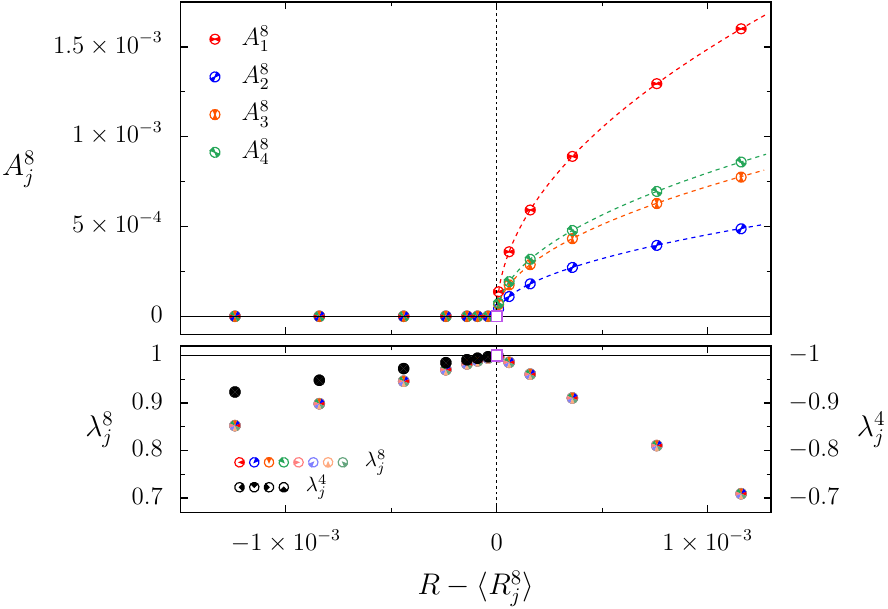}
  \end{tabular}
  \caption{Analysis of PD$_3$. (a) Torque amplitudes $A_j^8$ and (b) Multipliers $\lambda_j^8$ (coloured, sampling $\tau(\ell)$ with $N=8$) and $\lambda_j^4$ (black, sampling with $N=4$), as a function of inner cylinder Reynolds number $R-\langle R_j^8\rangle$. The square root fits are indicated with dashed lines. The lower bounds for the uncertainty of each point, shown as error-bars, are sufficiently small to be imperceptible. Following the graphical representation of the $\tau_j^8$ sequences of figure \ref{fig:PowFitPD8}, each of the amplitudes is represented with a double-sectored-disk that results from the superposition of the oppositely-oriented sectored-disks of the two torques that define the amplitude, hence the hourglass appearance of the symbols.
  }\label{fig:PD8}
\end{figure}
{Figure~\ref{fig:PD8}a indicates that there is a well-defined critical value of $R$, beyond which all four $A_j^8$ start growing following a square root law, as expected for a period doubling bifurcation.}
%
To accurately {pinpoint} the critical value of the parameter for PD$_n$, a fit of the form
\begin{equation}\label{eq:SqrFit}
  A_j^J(R) = a_j^J\sqrt{R-R_j^J}
\end{equation}
with $J=2^n$ {is} applied to the first few points after the period doubling bifurcation.
The fits provide, besides the $a_j^J$ values for the $J/2$ scaling factors, $J/2$ independent estimates $R_j^J$ for the critical value of the parameter.
A unique estimate for the critical value is then obtained by taking the arithmetic mean of the $J/2$ individual estimates as $R_n=\langle R_j^J\rangle$.
Performing the $J/2=4$ fits to the data points of figure \ref{fig:PD8}a (dashed lines), yields $R_3=\langle R_j^8\rangle=395.71184156\pm 5\times10^{-8}$ (violet square) for PD$_3$.

The eight $\lambda_j^8$ multipliers, shown in figure~\ref{fig:PD8}b, behave as expected across a period doubling bifurcation. The eight individual fits \eqref{eq:PowFit} to the eight individual $\tau_j^8$ subsequences produce nearly identical values of $\lambda_j^8$ at any given $R$, hence the multicoloured filled circles, resulting from the superposition of the eight oriented sectored disks. The unique estimate of the multiplier at any given $R$, obtained as the arithmetic mean $\lambda_8=\langle\lambda_j^8\rangle$, increases towards unity as the period doubling point is approached from either side. Since the solution above PD$_3$ is indeed a P$_8$, $\lambda_8$ is directly its multiplier. This is not so for the P$_4$ solution below PD$_3$. Plots analogous to those in figure~\ref{fig:PowFitPD8}a but sampling with $J=4$ show that convergence onto the four asymptotic branches follows an oscillatory pattern. The multiplier of the P$_4$ orbit in the vicinity of PD$_3$ must be either computed from fits to the torque subsequences sampled with $J=4$ or indirectly from $\lambda_8$ according to $\lambda_4=\langle\lambda_j^4\rangle=-\sqrt{\lambda_8}$. Either way, the multiplier of the stable P$_4$ in the vicinity of PD$_3$ is real, negative, and on the way of leaving the unit circle of the complex plane through $\lambda=-1$ as $R_3$ is approached from below (black disks in figure~\ref{fig:PD8}b), consistent with the period doubling nature of the bifurcation.

\section{Verification of Feigenbaum universality}\label{sec:FeiUni}

Signs that the period doubling cascade may exhibit some sort of universality are already apparent from figure~\ref{fig:PDcascade}, where all orbits computed up to period $N=128$ (past PD$_7$) are shown as a function of $R$.
\begin{figure}                                               
  \begin{center}
    \begin{tabular}{c}
      (a) \hspace{10em} (b) \hspace{10em} (c) \\
      \includegraphics[width=0.95\linewidth]{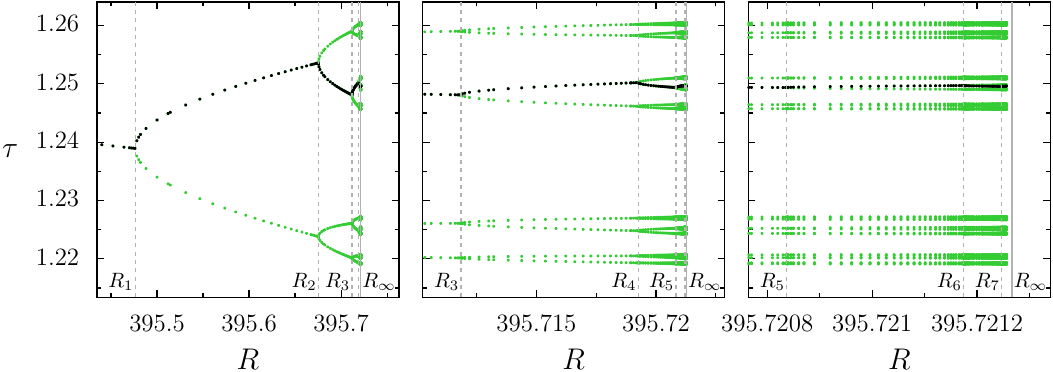}\\
    \end{tabular}
  \end{center}
  \caption{Successive magnifications of the period doubling cascade. {Same data as in figure~\ref{fig:biffigP1}b, truncated at the accumulation point $R_{\infty}$, the central branch indicated with black dots.} Overview (a) and two successive levels of magnification (b) and (c). The first few period-doubling bifurcations ($R_n$, dashed) and the accumulation point ($R_\infty$, solid) are indicated with grey lines and labels.
  }
\label{fig:PDcascade}
\end{figure}
Successive magnifications, presented in \ref{fig:PDcascade}b and \ref{fig:PDcascade}c, exhibit branch structures very similar to that of the full cascade of \ref{fig:PDcascade}a. Yet, at this stage, there is no guarantee that the universality observed here is of the same kind as discovered by Feigenbaum.

In order to ascertain Feigenbaum universality, we have computed the sequence of critical $R_n$ values along the period doubling cascade up to order $n=7$ to very high accuracy, using the method described in \S\ref{sec:determination}.
The results are summarised in table~\ref{tab:PD}.
\begin{table}
  \centering
  \begin{tabular}{c|c|r@{.}lr@{$\times$}l|r@{.}lr@{$\times$}l|r@{.}lr@{$\times$}l}
    $n$ & $N=2^n$ & \multicolumn{4}{c|}{$R_n$} & \multicolumn{4}{c|}{$\delta_n$} & \multicolumn{4}{c}{$\alpha_n$} \\
    \hline
    1 &   2 & 395&4762796241073  & $\pm6.9$&$10^{-13}$ & \multicolumn{2}{c}{-} & \multicolumn{2}{c|}{-} & \multicolumn{2}{c}{-} & \multicolumn{2}{c}{-} \\
    2 &   4 & 395&67542342   & $\pm5.3$&$10^{-7}$ & \multicolumn{2}{c}{-} & \multicolumn{2}{c|}{-} & \multicolumn{2}{c}{-} & \multicolumn{2}{c}{-} \\
    3 &   8 & 395&711841564  & $\pm4.2$&$10^{-8}$ & 5&46826 & $\pm1.0$&$10^{-4}$ & -2&6668339 & $\pm8.1$&$10^{-6}$ \\
    4 &  16 & 395&719257726  & $\pm8.7$&$10^{-8}$ & 4&91065 & $\pm1.6$&$10^{-4}$ & -2&611741 & $\pm1.6$&$10^{-5}$ \\
    5 &  32 & 395&72083697363& $\pm2.9$&$10^{-10}$ & 4&69601 & $\pm3.4$&$10^{-4}$& -2&496646 & $\pm2.9$&$10^{-5}$ \\
    6 &  64 & 395&7211746289 & $\pm4.7$&$10^{-9}$ & 4&67710 & $\pm3.3$&$10^{-4}$ & -2&510893 & $\pm2.8$&$10^{-5}$ \\
    7 & 128 & 395&7212469242 & $\pm1.4$&$10^{-9}$ & 4&67050 & $\pm4.6$&$10^{-4}$ & -2&500577 & $\pm4.4$&$10^{-5}$ \\
    $\dots$ & $\dots$ & \multicolumn{2}{c}{$\dots$} & \multicolumn{2}{c|}{$\dots$} & \multicolumn{2}{c}{$\dots$} & \multicolumn{2}{c|}{} & \multicolumn{2}{c}{$\dots$} \\
    $\infty$ & $\infty$ &  395&721266624 & $\pm1.1$&$10^{-8}$ & 
    \multicolumn{4}{l|}{4.66920160910299\ldots} & \multicolumn{4}{l}{-2.50290787509589\ldots}
  \end{tabular}
  \caption{
  Confirmation of Feigenbaum universality along the period doubling cascade.
  The values of the critical inner cylinder Reynolds number $R_n$ at the $n$th period doubling bifurcation point PD$_n$ are used to compute the $n$th approximation, $\delta_n$, to Feigenbaum's first constant, according to \eqref{eq:dndef}. Approximations, $\alpha_n$,  to the second constant, are computed from central branch torque values following \eqref{eq:FeigSecConst}.
  Lower bounds to the uncertainties in the $R_n$, $\delta_n$, $\alpha_n$ and $R_\infty$ parameters have been rigorously estimated from the covariance matrices of the various fits involved in the process, combined with standard error propagation theory (note that we have not taken into account the error of the fit (\ref{eq:PowFit}), as conducting a systematic study is difficult).  
  The last row of the table corresponds to the accumulation point $R_{\infty}$ estimated by (\ref{estaccu}), and the actual values of Feigenbaum constants, $\delta_{\rm F}$ and $\alpha_{\rm F}$.
}
\label{tab:PD}
\end{table}
{For every $n\ge3$, the table also lists} the ratios $\delta_n$, defined as
\begin{equation}
  \delta_n = \dfrac{R_{n-1}-R_{n-2}}{R_{n}-R_{n-1}}, 
  \label{eq:dndef}
\end{equation}
which provide the best approximation to Feigenbaum's first constant $\delta_{\rm F}$ that may be obtained from data up to the $n$th period doubling.
The trend of $\delta_n$ exhibits evident signs of conforming to Feigenbaum universality, that is, $\delta_n \rightarrow \delta_{\rm F}$ as $n\rightarrow \infty$. 
It is instructive to compare these results with those in table~\ref{table_univ}. Limiting the analysis to $n=4$, as customarily done in the fluid dynamics literature, does not guarantee convergence to even the second significant digit. Securing three digits precision is an arduous task that typically requires the accurate determination of at least $n=7$ period doubling bifurcation points.

The $R_n$ sequence is expected to converge onto the accumulation point, $R=R_\infty$. This point may be computed by assuming that all period doubling bifurcations of order higher than available are perfectly Feigenbaum-universal and, therefore, satisfy (\ref{eq:dndef}) with the left hand side replaced by $\delta_{\rm F}$.
However, this naive approach leads to systematic error.
To obtain a better estimate of $R_{\infty}$ and, at the same time, bound the error, we first note that
\begin{eqnarray}\label{eq:AccExp}
  R_{\infty} = R_n+\sum_{m=n+1}^{\infty}{\Delta R_m}, 
\end{eqnarray}
where $\Delta R_m\equiv R_m-R_{m-1}$ denotes the interval between the $(m-1)$ and $m$th period-doubling bifurcations. 
Since (\ref{eq:dndef}) can be written as $\delta_n=\Delta R_{n-1}/\Delta R_{n}$, any successive $m$th interval may be expressed recursively in terms of previous intervals as
\begin{equation}
  \Delta R_{m}=\delta_m^{-1}\Delta R_{m-1}=\delta_m^{-1}\delta_{m-1}^{-1}\Delta R_{m-2}=\dots=\delta_m^{-1}\delta_{m-1}^{-1}\cdots\delta_{n+1}^{-1}\Delta R_{n}=\Delta R_{n}\prod_{\ell=n+1}^{m}\delta_\ell^{-1}.
\end{equation}
The accumulation point can then be formally expressed in terms of the n$th$ period doubling and the immediately preceding interval as
\begin{eqnarray}\label{eq:AccExp2}
  R_{\infty} 
  = R_n+\Delta R_n\left[\sum_{m=n+1}^{\infty}{\prod_{\ell=n+1}^{m}\delta_\ell^{-1}}\right].
\end{eqnarray}
If $\delta_\ell=\tilde{\delta}_n>1$ were constant for all $ \ell>n$, then the terms in the square brackets would form a geometric series of sum $1/(\tilde{\delta}_n-1)$. We conveniently define the {\it average} ratio beyond PD$_n$ as
\begin{equation}\label{eq:deltatilde}
  \tilde{\delta}_n = 1 + \left[ \sum_{m=n+1}^{\infty}\displaystyle{\prod_{\ell=n+1}^{m}{\delta_\ell}^{-1}} \right]^{-1},
\end{equation}
so that (\ref{eq:AccExp2}) simplifies to
\begin{equation}\label{eq:Rinf}
  R_\infty = R_n + \dfrac{\Delta R_n}{\tilde{\delta}_n-1} = \dfrac{\tilde{\delta}_n R_n - R_{n-1}}{\tilde{\delta}_n-1}.
\end{equation}
The constant $\tilde{\delta}_n$ is unknown and may only be estimated. For $n$ sufficiently large, the sequence $\delta_\ell$ for all $\ell>n$ is expected to approach $\delta_{\rm F}$ monotonically from above. Under this assumption, inspection of (\ref{eq:deltatilde}) provides the inequalities $\delta_{\rm F}< \delta_{n+1}< \tilde{\delta}_n < \delta_n$. Accordingly, upper and lower bounds for $R_\infty$ can be obtained as
\begin{equation}\label{estaccu}
  R_\infty \in \left[R_n+\dfrac{R_n-R_{n-1}}{\delta_n-1},R_n+\dfrac{R_n-R_{n-1}}{\delta_{\rm F}-1}\right],
\end{equation}
by substituting $\tilde{\delta}_n=\delta_n$ and $\tilde{\delta}_n=\delta_{\rm F}$ in (\ref{eq:Rinf}), respectively.
Using PD$_7$, the upper and lower bounds for $R_\infty$ are already within the uncertainty range of one another (see table~\ref{tab:PD}) and, combined, provide an accurate estimate for the accumulation point $R_\infty=395.721266624\pm1.1\times10^{-8}$.

\begin{figure}
  \centering
   \includegraphics[width=.95\linewidth]{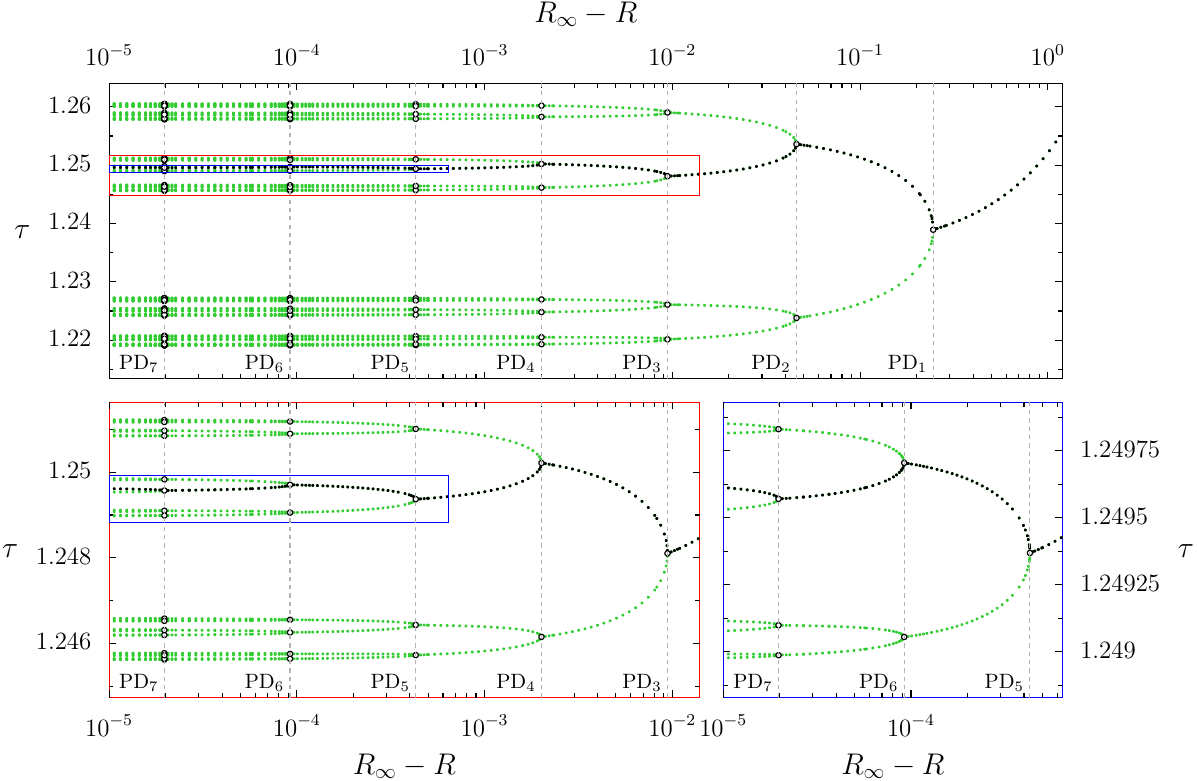}
  \caption{{Self-similarity of the period doubling cascade. Same data as figure~\ref{fig:PDcascade} in log scale, the central branch represented with black dots.}
  The coloured bottom panels are successive magnifications of the boxes in the top panel. The bifurcations are indicated with dashed vertical lines and labeled PD$_n$ according to their order.}\label{fig:PD}
\end{figure}

Following the determination of the accumulation point, we can now plot the period doubling cascade as a function of $R_\infty-R$ in logarithmic scale (figure~\ref{fig:PD}).
The asymptotic approach to universality and the accuracy with which the accumulation point has been obtained are evident from the similarities observed after two consecutive magnifications of the period-doubling cascade (red and blue bottom panels).

The occurrence of Feigenbaum universality implies that the dynamics of the Navier-Stokes system is governed by a (nearly) one-dimensional discrete map. Figure~\ref{fig:Racc}a depicts a three-dimensional phase map projection of the orbit at the accumulation point $R_\infty$.
\begin{figure}                                                                 
  \begin{center}
  \begin{tabular}{cc}
    (a) & (b) \\
      \includegraphics[height=.38\linewidth]{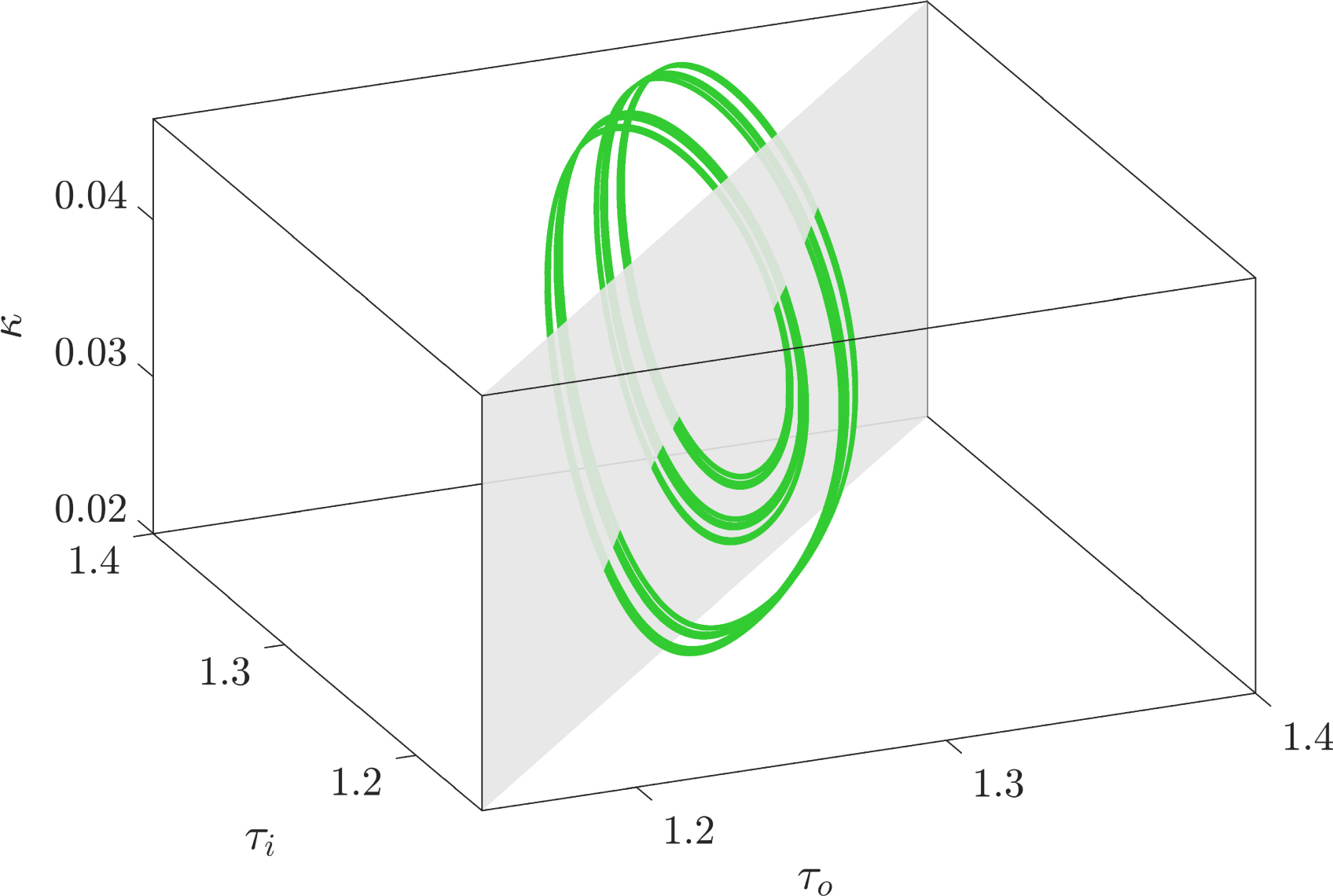} &
      \includegraphics[height=.38\linewidth]{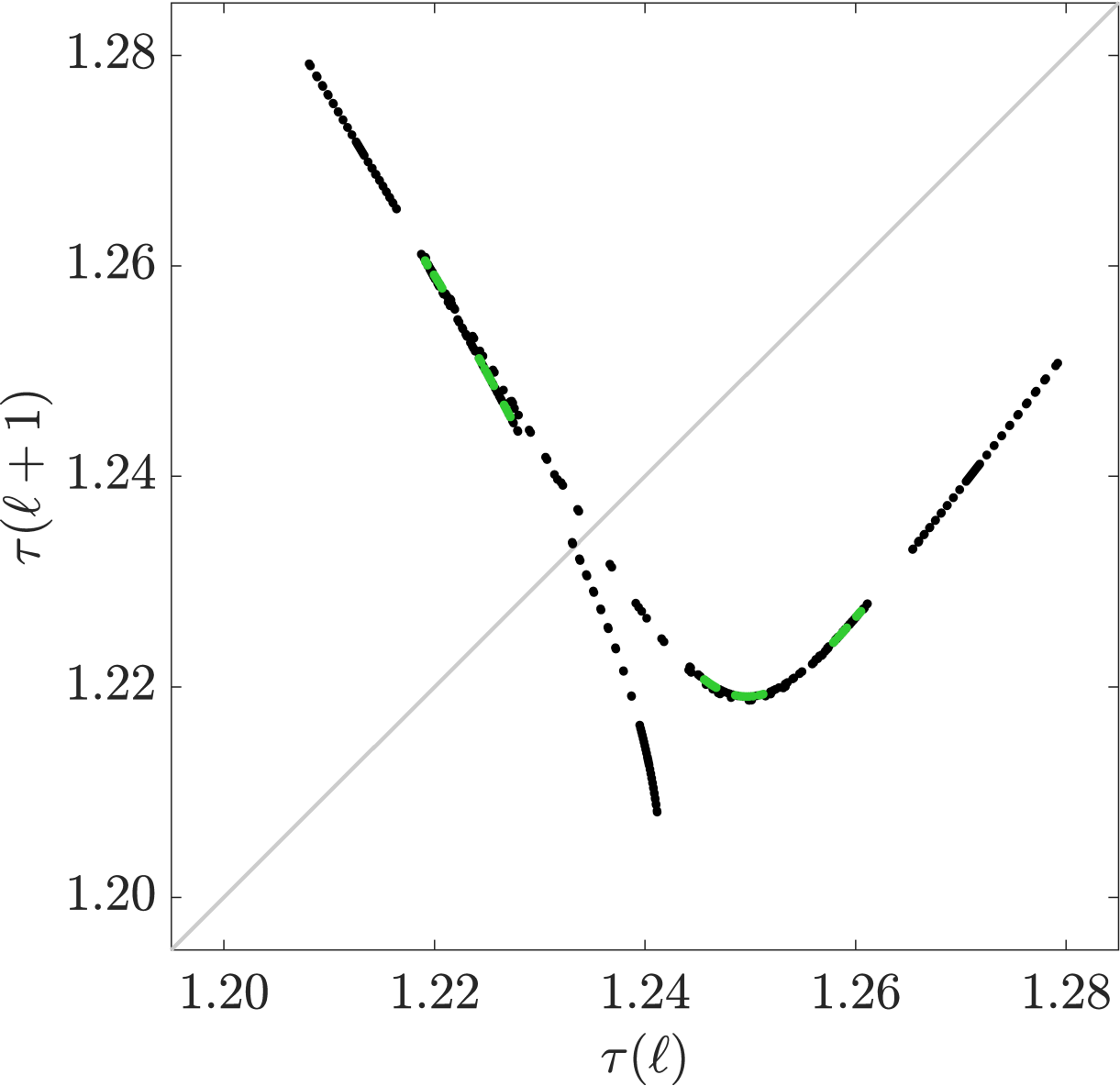} \\
   \end{tabular} 
 \end{center}  
  \caption{The {phase portrait} at $R_{\infty}$. (a) Three dimensional phase map projection of the chaotic attractor. (b) The first return iteration map on $\Sigma$ populated by the chaotic attractor (green dots) and transients leading back to it following a perturbation (black).
  }
\label{fig:Racc}          
\end{figure} 
If the manifold has dimension one, the coefficients $\mathbf{a}$ on {the Poincar\'e section} $\Sigma$ {introduced in \eqref{eq:PoincSec}} can be uniquely parameterised by the torque, such that the first return map $f$ defined by $\tau(\ell+1)=f(\tau(\ell))$ contains all the long-term properties of the Poincar\'e map. 
The aperiodicity of the orbit becomes apparent when plotting $\tau(\ell+1)$ as a function of $\tau(\ell)$ (green dots in figure~\ref{fig:Racc}b). The structure of the map on $\Sigma$ is, however, not clear, as the $\Sigma$-crossings of the orbit are sparsely distributed (see figure~\ref{fig:Racc}a). Perturbing the orbit, dropping the first initial transients, and plotting the remainder of the transients, conveniently assists in populating the map more densely (black points). {All the points thus obtained fall within a narrow, nearly one-dimensional band, {seemingly connected by} a smooth curve. This indicates that the discrete-time dynamics on $\Sigma$ in the vicinity of the accumulation point can be well approximated by a one-dimensional map $f$. Note that, in the range of $\tau$ for which $f$ appears to be multivalued (see the black points), the lower branch is never visited by the attractor and can, therefore, be ignored as regards its dynamics. Accordingly,} the function $f$, with its leftmost branch duly trimmed, becomes unimodal, and its smooth extremum (the minimum) admits a quadratic approximation. 
Note that unimodal maps with non-quadratic extrema provide universality constants that deviate from Feigenbaum's well-known values \citep{Bri91}. 


For period-doubling cascades in one-dimensional unimodal maps, approximations of the second Feigenbaum constant,
$\alpha_{\rm F}$, are often obtained from the {\it central branch} of the bifurcation diagram, which is generated by the periodic point that is nearest to the extremum of the map. {Since producing} accurate bifurcation diagrams of cascades undergone by fluid flow systems is an arduous task, \citet{Fe79} {used instead} the frequency spectrum, recorded at a fixed value of the parameter, of a natural convection experiment.
In the subsequent studies summarised in table \ref{table_univ}, no evaluation of agreement with $\alpha_{\rm F}$ was conducted.


On the other hand, we have a highly accurate bifurcation diagram available, rendering direct assessment of agreement with the second constant possible.
{In the torque sequences presented so far, we have indexed the $N$ distinct points visited by every periodic solution P$_N$ as $j=1,2,\dots N$ such that $j=1$ is the central branch point (the black dots in figure~\ref{fig:PD}).}
{Thus, using the torque values} of the central branch at PD$_m$, $\hat{\tau}_{m}=\tau_1^{2^m}(R_m)$, we can compute the ratios
\begin{eqnarray}\label{eq:FeigSecConst}
%
\alpha_n = \dfrac{\hat{\tau}_{n-1}-\hat{\tau}_{n-2}}{\hat{\tau}_{n}-\hat{\tau}_{n-1}}
\end{eqnarray}
to be compared against the $\alpha_{\rm F}$ appearing in the Feigenbaum-Cvitanovi\'c functional equation.
The values of $\alpha_n$ thus obtained from the bifurcation diagram, listed in table \ref{tab:PD} alongside $\delta_n$, evince gradual convergence toward Feigenbaum universality.

\section{Prediction of the period doubling route to chaos}\label{sec:UniMap}

{{Following} the seminal work of \citet{Fe79}, who {showcased} the first comparison with experimental results, \citet{Fe80,Fe82,Fe83} further extended renormalisation theory to include the universal scaling of all periodic points, not limited to the central branch, at different values of the parameter. The 
trajectory scaling function theory, which unravels the scaling between orbits featuring the same stability properties (i.e., having the same  multiplier), was later applied to interpret experimental data on natural convection by \citet{Belmonte-1988}.}

{The {occurrence} of universality {entails} that, for any given unimodal map, {one} should be able to predict the entire structure of the bifurcation diagram near the accumulation point using {data} from only the initial few period-doubling bifurcations. 
Unfortunately, the trajectory scaling function method does not provide an easy way to predict the bifurcation cascade.
Moreover, orbits become aperiodic beyond the accumulation point, 
{hindering direct comparison with orbits at different parameter values.}
{In this section, we} devise a method for predicting the unfolding of the bifurcation cascade up to and beyond the accumulation point.}

Consider a general one-dimensional discrete-time dynamical system described by 
\begin{eqnarray}
x_{\ell+1}=f_{a}(x_\ell).\label{eq:1dsystem}
\end{eqnarray} 
We assume that the function $f_{a}(x)$ is unimodal, and that a fixed point of the system \eqref{eq:1dsystem} undergoes a period doubling cascade as the parameter $a$ is increased. 
In drawing parallels with our Navier-Stokes problem, we will later identify $a$ and $f_a$ with the Reynolds number $R$ and the function approximating the return map of the torque sequence, respectively.
%
Our claim, motivated by \citet{Fe82}, is that the approximation
\begin{eqnarray}
\frac{f^{2^{n}}_{a}(x)-X}{\gamma} \approx  
 U^{2^n}_{M} \left ( \frac{x-X}{\gamma} \right), \qquad M= \frac{a-a_{\infty}}{\mu}\label{eq:approx}
\end{eqnarray}
holds as long as $2^n$ is sufficiently large, 
\begin{eqnarray}
\left |\frac{\delta^{n} (a-a_{\infty})}{\mu} \right |\ll 1, \qquad \left | \frac{\alpha^n(x-X)}{\gamma} \right | < 1,\label{eq:condition}
\end{eqnarray}
and provided the constants $\gamma$, $X$, $\mu$, and $a_{\infty}$ are suitably chosen.
The derivation of \eqref{eq:approx} and \eqref{eq:condition} is given in Appendix~\ref{app:UniMap}.
To simplify the notation, we use functional powers to denote repeated application of a function (e.g. $f^2_a(x)=f_a(f_a(x))$). Also, we have dropped the F subscript when referring to Feigenbaum's first ($\delta$) and second ($\alpha$) constants.

{{Here,} the function $U_M$ appearing in the approximation (\ref{eq:approx}) is defined in terms of the Feigenbaum function $G$ and the Feigenfunction $\varPhi$ (with the normalisation $\varPhi(0)=1$) as
\begin{eqnarray} 
U_M(\xi) = G(\xi)+M\varPhi(\xi).\label{eq:UMMM}
\end{eqnarray}
The $\delta$ and $\varPhi$ satisfy the eigenvalue problem $\mathcal{L}_G[\varPhi]=\delta \varPhi $. The precise form of the linear operator $\mathcal{L}_G$ is given in Appendix~\ref{app:UniMap}.
The nature of $G$ and $\varPhi$ is well understood \citep{CoEck80,Bri91,ThurlbyPhD2021}. For our purposes, it will suffice to know that they are both smooth even functions, and that numerical recipes {for their computation abound in the literature}.}

If the four (map-dependent) parameters, $\gamma$, $X$, $\mu$, and $a_{\infty}$, can be computed from the first few period doublings, then \eqref{eq:approx} provides a useful tool for predicting the behaviour of the dynamical system \eqref{eq:1dsystem}, subject to the constraints outlined by \eqref{eq:condition}. In practice, fixing $n=2$ and using data up to PD$_{n+2}$ (i.e. PD$_4$) yields already reasonably accurate results. The procedure for prediction is straightforward, as we shall see shortly. All it takes is extracting {{\it template} branches}  from the bifurcation diagram of the {one-dimesnional} dynamical system
\begin{eqnarray}
\xi_{\ell+1}=U_M(\xi_\ell),\label{eq:1dsystemxi}
\end{eqnarray}
and then stretch and translate {them to align} with {corresponding branches of} the bifurcation diagram of \eqref{eq:1dsystem}. The constants $X$ and $\gamma$ translate and scale the state $x$, while $a_\infty$ and $\mu$ do the same for the parameter $a$, respectively. The constant $a_{\infty}$ corresponds to the predicted accumulation point of the dynamical system \eqref{eq:1dsystem}. Thus the first condition in \eqref{eq:condition} requires that the parameter $a$ be sufficiently close to the accumulation point. Likewise, $X$ corresponds to the predicted position of the extremum of the map $f_a(x)$ at $a=a_{\infty}$, and hence the second condition in \eqref{eq:condition} limits the validity of the approximation to a close neighbourhood of the extremum of the map $f_a(x)$, i.e., the central branch of the cascade. In brief, the conditions \eqref{eq:condition} imply that universality applies to the bifurcation diagram only locally. We will see later how the prediction can, nevertheless, be extended to the rest of branches.

The bifurcation diagram of the dynamical system \eqref{eq:1dsystemxi} is depicted in figure~\ref{fig:univ}a.
\begin{figure}
  \centering
  \begin{tabular}{cc}
  (a) & (b)\\
  \includegraphics[height=.4\linewidth]{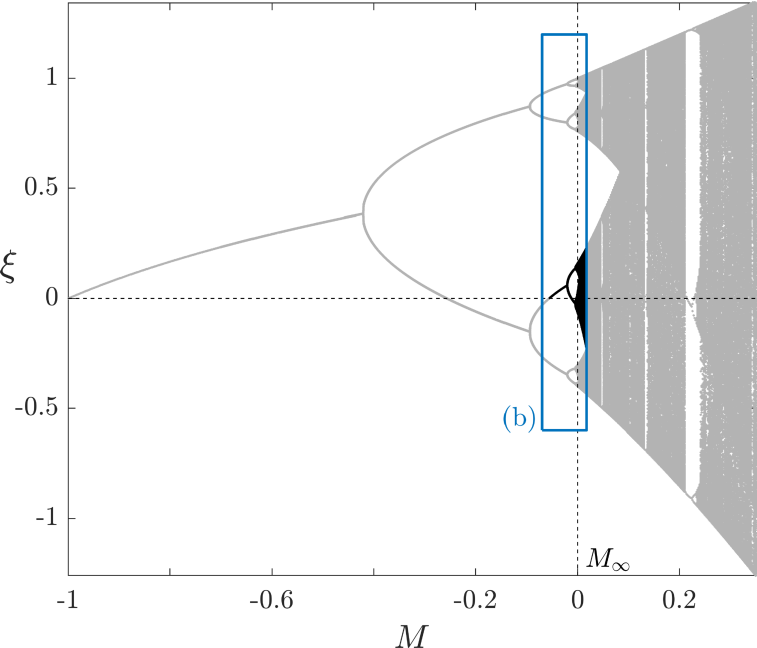} & \includegraphics[height=.4\linewidth]{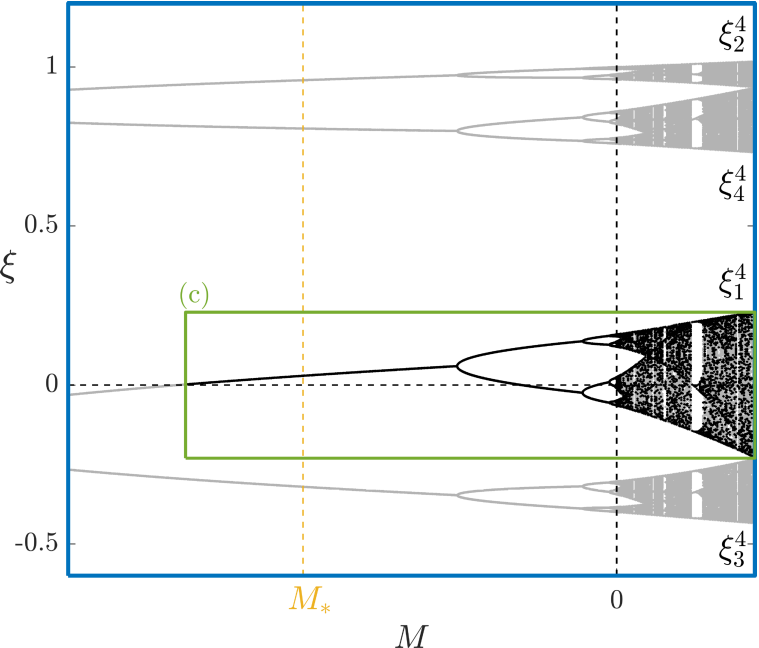}\\
  (c) & (d)\\
  \includegraphics[height=.4\linewidth]{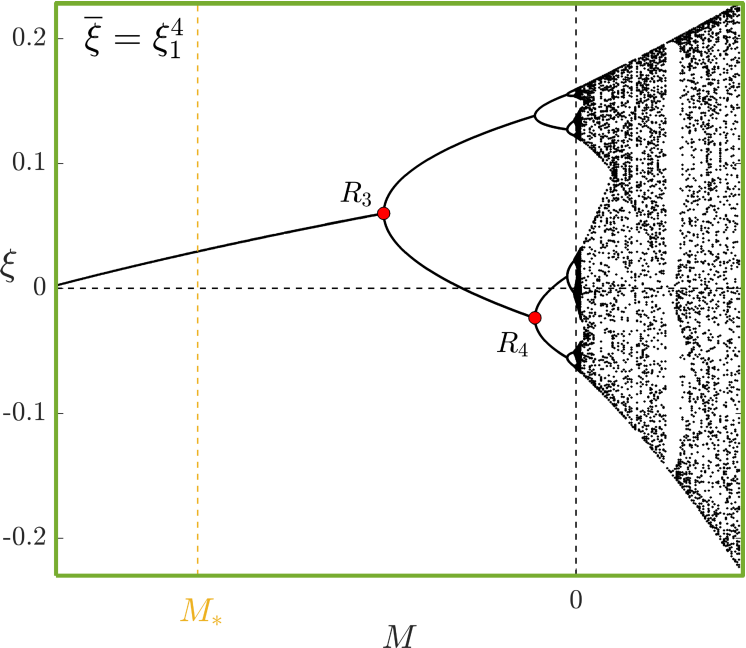} & \includegraphics[height=.4\linewidth]{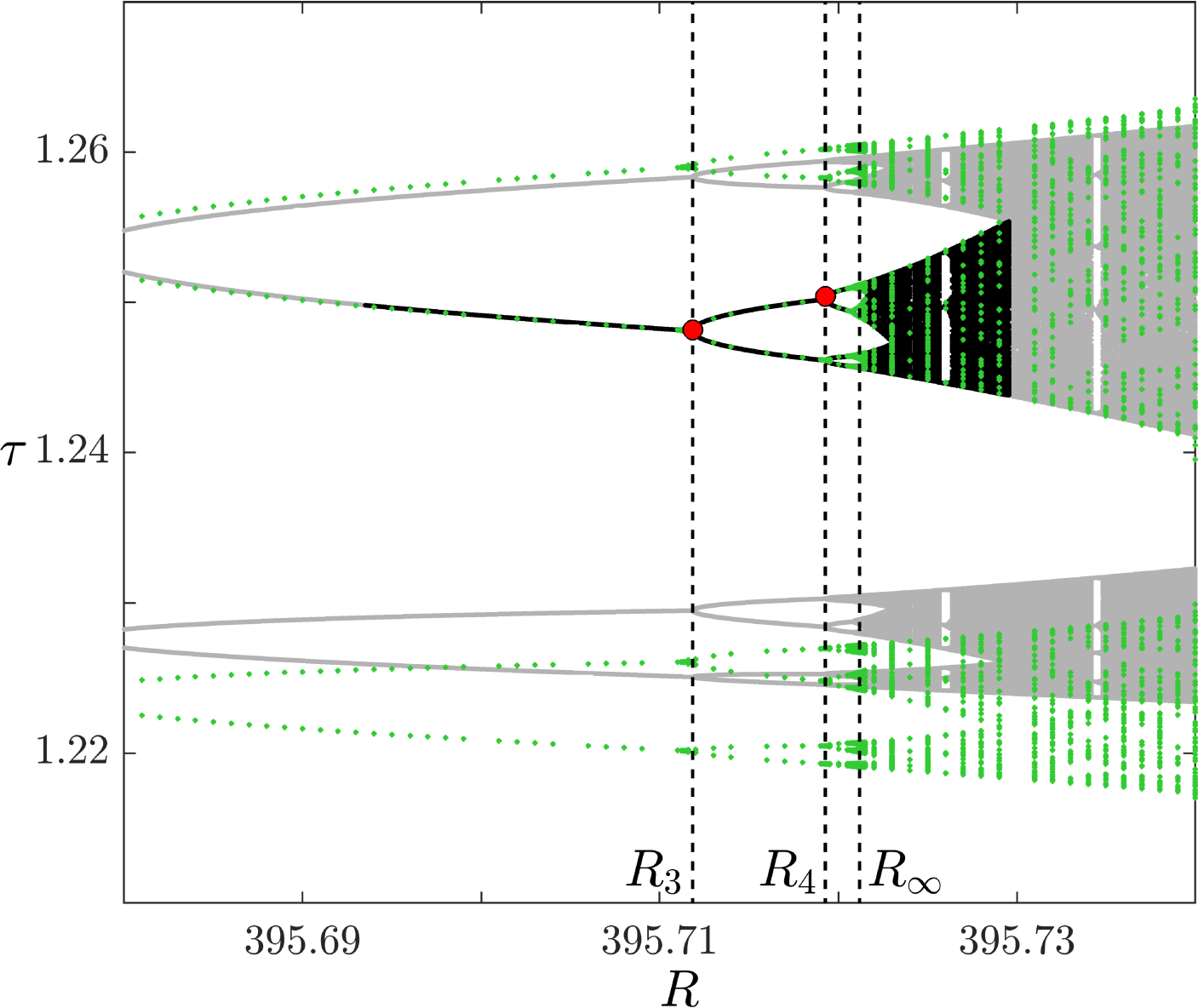}
  \end{tabular}
  \caption{
    Bifurcation diagram of the one-dimensional map (\ref{eq:1dsystemxi}).
    (a) The full period doubling cascade. (b) Magnification around the accumulation point $M_\infty$. The $2^n=4$ distinct sets of points arising for $M>M_*$ are labeled as $\xi_j^n$, $j=1,2,3,4$. Black dots are used for the central branch, grey for the rest. (c) Close-up of the central branch $\overline{\xi}=\xi_1^n$. {The red bullets at PD$_3$ and PD$_4$ are the points $(M,\xi)=(M_{n+1},\hat{\xi}_{n+1})$ and $(M_{n+2},\hat{\xi}_{n+2})$ selected for the matching with DNS data. }
    {(d) The full cascade transformed by (\ref{eq:simpletrans}) with $(\gamma,X,\mu,a_{\infty})$ obtained by (\ref{eq:four}), compared with the period doubling cascade in Taylor-Couette flow (green dots). 
}
  }\label{fig:univ}
\end{figure}
It is easy to show that, for $M=-1$, the system has a stable fixed point at the origin ($\xi=0$), and that $M=M_\infty=0$ is the accumulation point of the period-doubling cascade. {Assume that} fixing the parameter $M$ to a small (in modulus) negative value
produces a stable orbit of, say, period $N=2^n$ with $n\in\mathbb{N}$. An example of $M=M_*$ producing a period-4 orbit ($n=2$) is shown in figure~\ref{fig:univ}b, which contains a magnification of figure~\ref{fig:univ}a around the accumulation point. Each of the $2^n$ branches at $M=M_*$ {initiates} its own period doubling cascade as $M$ is increased from $M_*$ towards $M_\infty$. After the accumulation point, the $2^n$ distinct clouds of points induced by the $2^n$ separate branches start merging in pairs {following a} reverse cascade until eventually forming one sole object. We have labeled the four cascades as $\xi_j^4$ ($j=1,2,3,4$) in figure~\ref{fig:univ}b and shown them up to the point where they first merge in two pairs ($\xi_1^4$ with $\xi_3^4$ and $\xi_2^4$ with $\xi_4^4$). 
{We shall see that each {of these} cascade branch{es} {can} act as a template in our prediction, {but first, let us} focus on the cascade induced by the central branch, {henceforth} denoted as $\overline{\xi}=\xi_1^4$ (see figure~\ref{fig:univ}c).}

To illustrate how the approximation \eqref{eq:approx} operates, {we} begin by fixing $M=M_*$ in figure~\ref{fig:univ}b, where we have an orbit of period $N=2^n=4$, so $n=2$. {Each template $\xi_j^4$ corresponds to a single point at this {value of the} parameter, {with the $j$ labels following the same sampling procedure} of (\ref{eq:kNj}) with $J=2^2$.} Assuming that \eqref{eq:approx} is sufficiently accurate, we should be able to compare system (\ref{eq:1dsystemxi}) at $M=M_*$ with system (\ref{eq:1dsystem}) at $a_*=\mu M_*+a_{\infty}$. At $M=M_*$, the template solution $\overline{\xi}_{*}=\overline{\xi}(M_*)$ satisfies $U_{M_*}^{2^n}(\overline{\xi}_*)=\overline{\xi}_*$. Substituting $x=\overline{x}(a_*)=\overline{x}_*=\gamma \overline{\xi}_*+X$ into \eqref{eq:approx} yields $f_{a_*}^{2^n}(\overline{x}_*)\approx \overline{x}_*$, which implies that system \eqref{eq:1dsystem} should exhibit a period-$2^n$ point well approximated by $x=\overline{x}_*$ at $a_*=\mu M_*+a_{\infty}$.
This orbit {should} be stable because differentiation of \eqref{eq:approx} using $\xi=(x-X)/\gamma$ demands that
\begin{eqnarray}\label{eq:diff}
\left.\dfrac{d}{dx}\left(f^{2^{n}}_{a_*}\right)\right|_{x=\overline{x}_*}\approx \left.\dfrac{d}{d\xi}\left(U^{2^n}_{M_*}\right)\right|_{\xi=\overline{\xi}_*},
\end{eqnarray}
which tells that the multiplier of the $2^n$ orbit of the map $f_a$ must be reasonably well approximated by that of the universal map $U_M$. 
This result is actually not surprising because \eqref{eq:approx} {indicates} that the dynamics induced by the map $f^{2^n}_a(x)$ around $x=X$ are quantitatively similar to those induced by $U_M^{2^n}(\xi)$ around $\xi=0$. 
{
The same argument can be made for the correspondence between solutions $\overline{\xi}(M)$ and $\overline{x}(a)$ at any $a=\mu M+a_\infty$, which need no longer be a single point as the central branch cascade unfolds. For $a\in (a_*,a_{\infty})$, stable periodic orbits in the two systems can be related in the same way, and the approximate equality of multipliers follows from (\ref{eq:diff}) analogously. Moreover, the dynamical similarity is not limited to periodic orbits but applies also to aperiodic solutions {for $a>a_\infty$}. In general, we must interpret that the template solution $\overline{\xi}=\overline{\xi}(M)$ consists of a set of points satisfying $U_{M}^{2^n}(\xi_i)\in \overline{\xi}$ for all $\xi_i \in \overline{\xi}$ (i.e. the set is invariant under $U_{M}^{2^n}$).
Then \eqref{eq:approx} implies that the set $\overline{x}(a)$, defined as the collection of points $x_i=\gamma \xi_i+X$ for every $\xi_i \in \overline{\xi}(M)$, is invariant under $f_{a}^{2^n}$. 
Chaotic solutions {related by $a=\mu M+a_\infty$} will have approximately the same Lyapunov exponents due to (\ref{eq:diff}).
All in all, the similarity of central branch dynamics between the two systems extends naturally all the way up to the accumulation point and beyond. 
}

In order to find an adequate set of values for $\gamma, X, \mu,$ and $a_{\infty}$, one must pick two points in one of the bifurcation diagrams and find their analogues in the other.
As a proof of concept, we select the first and second period doubling bifurcation points that are encountered upon increasing $M$ beyond $M_*$, where we have the stable period-$2^n$ orbit, namely PD$_{n+1}$ and PD$_{n+2}$. We call the values of the parameter and of the central branch state variable $(M,\xi)=(M_{n+1},\hat{\xi}_{n+1})$ and $(M_{n+2},\hat{\xi}_{n+2})$ for PD$_{n+1}$ and PD$_{n+2}$, respectively (see figure~\ref{fig:univ}c). These same period doubling bifurcation points are found to occur at $(a,x)=(a_{n+1},\hat{x}_{n+1})$ and $(a_{n+2},\hat{x}_{n+2})$ for the dynamical system \eqref{eq:1dsystem}, which, in the context of our DNS computations, correspond to $(R,\tau)=(R_{n+1},\hat{\tau}_{n+1})$ and $(R_{n+2},\hat{\tau}_{n+2})$ of \S\ref{sec:FeiUni}. The matching of {both} points from one bifurcation diagram to the other is accomplished by enforcing $M_{m}= (a_{m}-a_{\infty})/\mu$ and $\hat{\xi}_{m}=(\hat{x}_{m}-X)/\gamma$ for $m=n+1$ and $n+2$. Solving the four resulting algebraic equations, we get 
\begin{equation}\label{eq:four}
  \begin{array}{rcl}    
    \gamma & = & \displaystyle{\frac{\hat{x}_{n+2}-\hat{x}_{n+1}}{\hat{\xi}_{n+2}-\hat{\xi}_{n+1}}},\\[1em]
    X & = & \displaystyle{\frac{\hat{\xi}_{n+2}\hat{x}_{n+1} -\hat{\xi}_{n+1} \hat{x}_{n+2}}{\hat{\xi}_{n+2}-\hat{\xi}_{n+1}}},\\[1em]
    \mu & = & \displaystyle{\frac{a_{n+2}-a_{n+1}}{M_{n+2}-M_{n+1}} }, \\[1em]
    a_{\infty} & = & \displaystyle{\frac{M_{n+2} a_{n+1} -M_{n+1}a_{n+2} }{M_{n+2}-M_{n+1}}}.
  \end{array}
\end{equation}
If we set $n=2$ and exploit the DNS results of figure~\ref{fig:PD}, the quadruplet of matching parameters takes the values 
{$(\gamma, X,\mu, a_{\infty})\approx (-0.02518,0.2496,-0.4629,395.7213)$.}
Finally, applying the transformation 
\begin{equation}\label{eq:simpletrans}
\overline{x}(a)=\gamma \overline{\xi}(M)+X, \qquad a=\mu M+a_{\infty}
\end{equation}
{to the full cascade {generated} by system \eqref{eq:1dsystemxi}} results in the branch arrangement shown in figure~\ref{fig:univ}d. The transformed cascade exhibits a remarkably good agreement with DNS for the central branch (black dots). {However, the quality of the match does not carry over to the rest of template branches. It is {precisely} in this sense that the approximation (\ref{eq:approx}) is to be considered only local.}

The prediction can, however, be straightforwardly extended to all $2^n$ 
branches of the cascade undergone by the system \eqref{eq:1dsystem}, by simply allowing for a different set of scaling and translation parameters for each of the branches. 
The transformation of the $j$th branch is performed according to
\begin{equation}\label{eq:branchbybranch}
x_j^J(a)=\gamma_j \xi_j^J(M)+X_j, \qquad a=\mu M+a_{\infty}, \qquad j=1,2,3,\dots,J,
\end{equation}
where the pairs $(\gamma_j,X_j)$ are obtained by matching the bifurcation points in the respective bifurcation diagrams, following the same procedure as for the central branch. 
The prediction thus obtained for the complete unfolding of the bifurcation cascade undergone by the DNS is shown in figure~\ref{fig:univ2}.
\begin{figure}
  \centering
  \includegraphics[width=.75\linewidth]{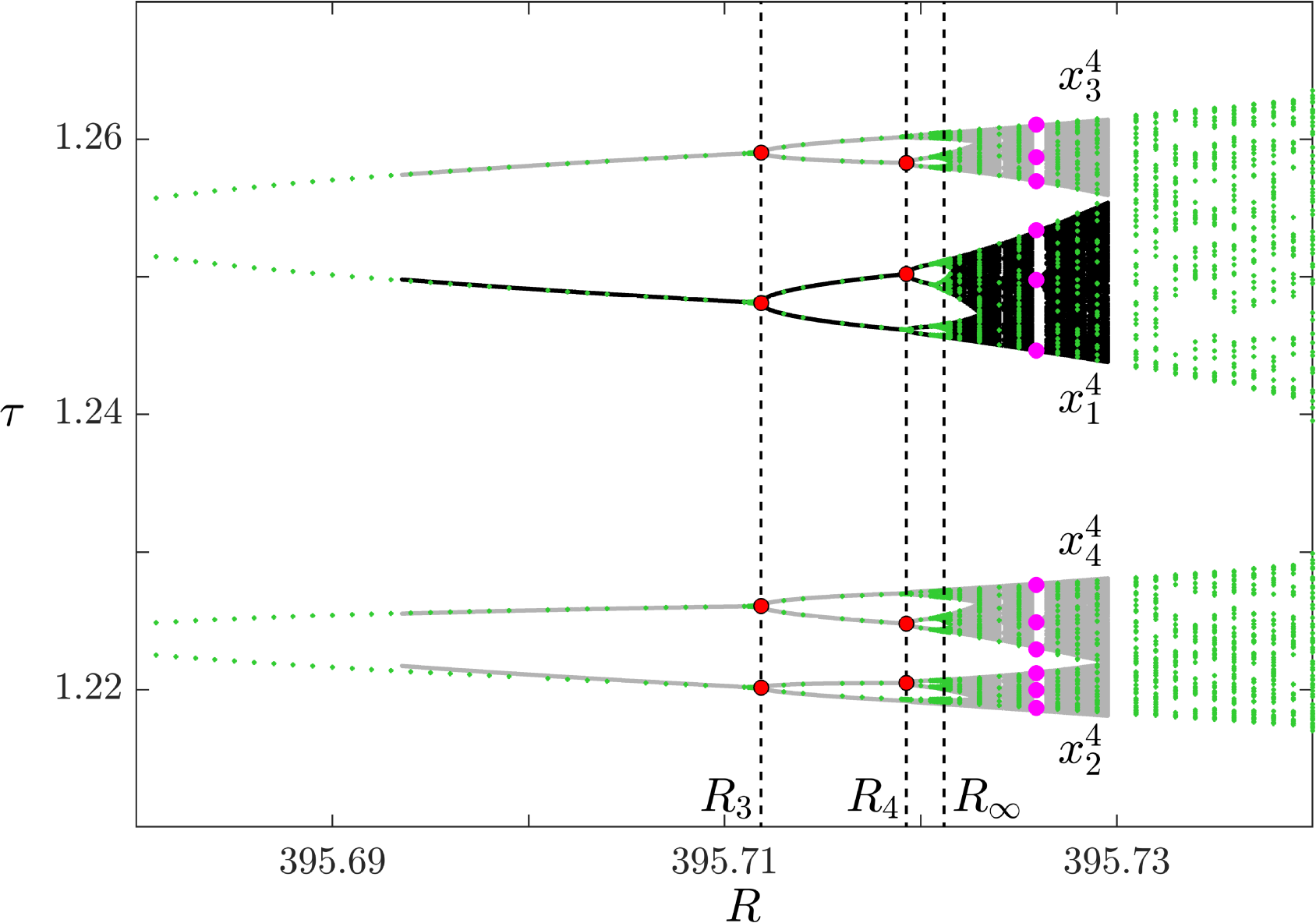}
  \caption{Branch-by-branch matching of period doubling cascades.
The black and grey points are the template branches shown in figure \ref{fig:univ}b, re-adjusted according to the rescaling (\ref{eq:branchbybranch}). The green points are the same DNS data as in figure \ref{fig:biffigP1}b. The matching process is based on the 8 red points. 
  The magenta points correspond to the P$_{12}$ orbit in the Taylor-Couette system.
  }
  \label{fig:univ2}
\end{figure}
{The agreement is now excellent for all 4 cascade branches. The success of the branch-by-branch matching process {follows from the fact that the map $f^{2^n}_a(x)$ has $J=2^n$ extrema, the approximation \eqref{eq:approx} being applicable to each individually.} {Note} that, in the chaotic regime after the accumulation point, the dynamics of the two systems are not expected to match in detail due to the sensitivity to initial conditions.
Nonetheless, the statistical properties should, in principle, be in fair agreement, as evinced by figure~\ref{fig:univ2}.} 

{
{For simplicity, we have limited our comparison across systems to orbits of the same period.} It is also possible to {relate} stable orbits of different periods, as \citet{Fe82} did, to demonstrate the self-similarity of the bifurcation diagram (see Appendix~\ref{app:UniMap}). Furthermore, our theoretical result is consistent with the fact that self-similarity also holds beyond the accumulation point, and that the Lyapunov exponent exhibits the scaling law observed by \cite{HuRu80}. However, a detailed comparison of statistical quantities such as Lyapunov exponents is beyond the scope of our analysis, given that statistical convergence requires unaffordably long simulation times in the vicinity of the accumulation point (see \cite{TiTaBe09}, for example). Instead, to gauge the accuracy of the prediction beyond the accumulation point, we have run DNS at a value of $R$ for which the occurrence of a periodic window with a P$_{12}$ orbit is anticipated. The time series of the Taylor-Couette system indeed appears to converge to a stable periodic orbit, which we have confirmed to be the Navier-Stokes counterpart of P$_{12}$ employing the PNK method (magenta dots in figure~\ref{fig:univ2}).}

{
All self-similar features of the template period-doubling cascade should also show up in any cascade generated by a unimodal map. This implies that any other unimodal map, including the long-studied logistic map, could have been employed as template. The advantage of the $U_M$ defined in (\ref{eq:UMMM}) stems from its natural emergence in the derivation of the approximation theory (as seen in Appendix~\ref{app:UniMap}), {which grants} very rapid convergence to universal dynamics {by construction}.} {Extending the prediction method to more general cases is straightforward using the results by \citep{Bri91}}. 

{
The core of our prediction method lies in completely separating the map-dependent scaling{/translating} parameters $(\mu, a_{\infty}, \gamma_j, X_j)$ from {all universality-related considerations}. A careful reading of Feigenbaum's work suggests that he was likely aware of this possibility. Notably, the method of trajectory scaling functions was specifically designed to eliminate all ingredients that depend on any particular map. However, to the authors' knowledge, there is no direct assertion in the literature pointing at the feasibility of such predictions, {let alone} their applicability beyond the accumulation point.}

\section{Conclusions}\label{sec:conclusions}

We have studied in detail a period-doubling cascade {that arises in} subcritical counter-rotating Taylor-Couette flow employing small computational domains of annular-parallelogram shape. The cascade {is seeded from a} family of drifting rotating waves discovered by W22 and eventually leads to a chaotic {regime fueled by the} self-sustained {process of wall-bounded shear flows}.
Although the Navier-Stokes equations should formally be considered an infinite-dimensional dynamical system, the dynamics are often confined to finite-dimensional manifolds in phase space, sometimes designated as inertial manifolds \citep{Te89}. This low dimensionality has recently attracted significant attention \citep{Di16,HaKaLiAx23}, as it affords valuable insight into the underlying dynamics of the flow, sometimes leading to simplified models whose numerical simulation is computationally more tractable. Our results can be interpreted as an extreme case {of simplification that ultimately leads} to a one-dimensional map representation. 

A pivotal outcome of {our} analysis is the confirmation of Feigenbaum universality to an unprecedented degree of accuracy in the context of fluid dynamics. Our success is largely ascribable to the fast development of computational power over the past few decades, along with the outstanding numerical accuracy of the methods we employ to solve Navier-Stokes flows, which altogether has rendered thorough parameter sweeps feasible. That said, establishing universality has required very long and costly numerical simulations, a refined parametric exploration, and the deployment of very accurate numerical techniques for the detailed analysis of the asymptotic convergence of relative periodic orbits, the assessment of their stability and the computation of period-doubling bifurcation points. Determining both the first and second Feigenbaum constants to the third significant digit demands the precise computation of up to the seventh period-doubling bifurcation along the cascade. This has enabled an accurate estimation of the accumulation point and, with it, the unambiguous exposure of the self-similar structure of the period-doubling bifurcation cascade. A key element to make the analysis systematic has been the deployment of a convenient Poincar\'e section based on torque balance, 
which provides a robust sampling method to validate theories of unimodal discrete-time dynamical systems. 

Furthermore, our results constitute the first confirmation of universality in a fluid flow problem subject to subcritical turbulent transition.
A detailed study of period doubling cascades is contingent on the existence and identification of a stable solution.  In subcritical shear flows stable solutions are rare and can only occasionally be found by working with minimal flow units and/or symmetry subspaces. 
 We were fortunate enough to find one such solution in W22 that happens to be at the origin of a period-doubling cascade.
}
It is also noteworthy that all solutions along the period doubling cascade exhibit spatial drift, a property that is unique to our system among the many previous studies on the subject (see table~\ref{table_univ}). As expected, the conventional Feigenbaum theory {becomes applicable to the Taylor-Couette} phase space once reduced by the method of slices \citep{BuCvDaSi15}.

Drawing from Feigenbaum's theory, we have further developed a method to predict the bifurcation diagram of the period-doubling cascade all the way up to the accumulation point and, remarkably, also beyond. This approach has broader applicability than the method of trajectory scaling functions \citep{Fe83,Belmonte-1988}, which is limited to scrutinising stable orbits with the same value of the multiplier. Our prediction method is simple: it only requires {stretching} and shifting a template bifurcation diagram produced by any {smooth unimodal map} {at hand}. 
The scaling and translating parameters must be chosen such that the template bifurcation diagram aligns with the period-doubling cascade of the target system (in our case, Taylor-Couette flow). This can be accomplished, for example, by matching two bifurcation points across systems. The remainder of the {\it transformed} template diagram provides the prediction for the target system. The prediction results are useful, among other things, to {anticipate the location of periodic} windows in the bifurcation diagram {of the target system} beyond the accumulation point. The emergence of a stable orbit in the Taylor-Couette system precisely within the expected parameter range attests to the exceptional predictive capability of the method.

Our analysis further provides a theoretical explanation for the self-similarity that occurs in the reverse cascade following the accumulation point (see the appendix). Since the theoretical result holds universally for unimodal maps, the excellent agreement between our DNS data and the reference one-dimensional map implies that self-similarity can be reasonably expected to extend beyond the accumulation point also for the Navier-Stokes system.

The observation of Feigenbaum universality in our Taylor-Couette setup provides evidence that the dynamics around the accumulation point can indeed be approximated by a nearly-one-dimensional discrete map when analysed on an appropriate Poincar\'e section. 
Our finding further provides a handy playground for testing the application of iconic theorems from the early days of chaos theory, such as Li-Yorke or Sharkovsky's \citep{Sh64,LiYo75}, as well as cycle expansion theory \citep{CAMTV16}, to a fluid dynamics problem. It is our intention to explore the significance and implications of these theories to the field of fluid dynamics in the near future.

{As briefly commented in section\,\ref{sec:introduction},}
{understanding the laminar-turbulent pattern-formation aspects of subcritical transition using stochastic theories \citep{Hof23DP} requires the prior occurrence of chaotic dynamics in the system.}
{In shear flows, these incipient chaotic dynamics are only transient, such that the underlying {\it chaotic saddle} is not accessible through experiments or DNS.}
The scenario we have dissected offers a possible mechanism for the formation of one such chaotic saddle because our period doubling cascade (i) occurs globally in the full annulus, and (ii) is unstable to subharmonic (spatially modulational) perturbations, as can be inferred from W22.
{Both the full and elongated domains of figure~\ref{fig:tcf1}b are simultaneously compatible with spiral turbulence and with the period doubling cascade we address here.}
Evidence for (ii) can {indeed} be {drawn from} W22, where the subharmonic instability of DRW, along with all solutions bifurcated from it in our small domain, was shown to contribute decisively to the formation of the laminar-turbulent helical pattern that is characteristic of the spiral turbulence regime. Therefore, although other mechanisms have been shown to produce localised chaotic sets that may be held responsible for intermittency in some cases \citep[see][for an example in channel flow]{PaYaDu23}, 
also a period doubling cascade might work concurrently with a modulational instability to the same effect.

\backsection[Acknowledgements]{
This research is supported by the Australian Research Council Discovery Project DP230102188 and the Ministerio de Ciencia, Innovación y Universidades (Agencia Estatal de Investigación, project nos. PID2020-114043GB-I00 (MCIN/AEI/10.13039/501100011033) and PID2023-150029NB-I00 (MCIN/AEI/10.13039 /501100011033/FEDER, UE). BW's and RA's research has been funded by the European Union’s Horizon 2020 research and innovation programme (Marie Sk\l{}odowska-Curie Grant Agreement No. 101034413). RA has also been funded by the Austrian Science Fund (FWF) 10.55776/ESP1481224.}

\backsection[Declaration of Interests]{
The authors report no conflict of interest.
}

\backsection[Author ORCIDs]{B. Wang, https://orcid.org/0000-0002-6229-0336; R. Ayats, https://orcid.org/0000-0001-6572-0621; K. Deguchi, https://orcid.org/0000-0002-3709-3242; A. Meseguer, https://orcid.org/0000-0002-2022-2001; F. Mellibovsky, https://orcid.org/0000-0003-0497-9052}

\appendix


\section{Derivation of a universal approximation to a unimodal map and the conditions for validity}\label{app:UniMap}

Consider the one-dimensional map \eqref{eq:1dsystem}, with $f_a(x)$ a unimodal function, undergoing a period-doubling cascade as the parameter $a$ is increased. The location where the function has a maximum for $a=a_\infty$, the accumulation point, is denoted as $X=\text{argmax} f_{a_\infty}(x)$. It will be convenient to shift the coordinate according to $y=x-X$ and define the shifted map as $F_{a}(y)=f_{a}(y+X)-X$. 

If $a$ is close to the accumulation point $a_{\infty}$, the function $F_{a}$ may be approximated by a Taylor expansion truncated at first order, i.e.
\begin{eqnarray}
F_{a}(y) \approx \varPsi (y)+(a-a_{\infty})\psi(y), \label{eqA:approxF}
\end{eqnarray}
where $\varPsi=F_a |_{a=a_{\infty}}$ and $\psi=\left. \partial_a F_{a}\right |_{a=a_{\infty}}$. Unimodal maps are known to be infinitely renormalisable at accumulation points \citep[see][for example]{LanIII82}. Therefore, for large $n$ and $a=a_\infty$, one must be able to find a function $g$, satisfying $\mathcal{R}[g]=g$, such that
\begin{eqnarray}
\mathcal{R}^n[\varPsi](y)\approx g(y).\label{eqA:Gapprox}
\end{eqnarray}
Note that $\gamma=g(0)$ is not necessarily unity and depends on the map $f_a$. 
The function $g$ and the Feigenbaum function $G$ introduced in \S\ref{sec:introduction} are related by 
\begin{eqnarray}\label{GGGG}
G(\xi)=g(\gamma \xi)/\gamma.
\end{eqnarray}

Next we consider the Fr\'echet derivative of $\mathcal{R}$, in the direction of function $h(y)$, and evaluated at function $f(y)$: 
\begin{eqnarray}
\mathcal{L}_{f}[h](y)&=&\lim_{\epsilon \rightarrow 0} \frac{\mathcal{R}[f+\epsilon h](y)-\mathcal{R}[f](y)}{\epsilon} = \alpha h\left (f \left (\frac{y}{\alpha}\right )\right )+\alpha f' \left (f \left (\frac{y}{\alpha} \right ) \right ) h\left (\frac{y}{\alpha} \right ). \label{eqA:Jdef}
\end{eqnarray}
The operative expression of $\mathcal{L}_{f}[h](y)$ in terms of $f$ and $h$ is a well-known result \citep[see][for example]{Bri91,ThurlbyPhD2021}.
The definition \eqref{eqA:Jdef}, combined with the expansion \eqref{eqA:approxF}, implies $\mathcal{L}_{\varPsi}[\psi] \approx (\mathcal{R}[F_{a}]-\mathcal{R}[\varPsi])/(a-a_{\infty})$, such that simple algebraic manipulation yields
\begin{eqnarray}
\mathcal{R}[F_{a}]\approx \mathcal{R}[\varPsi]+(a-a_{\infty})\mathcal{L}_{\varPsi}[\psi].\label{eqA:RFRF}
\end{eqnarray}
Moreover, it can be shown by induction that the $n$th renormalisation of $F_a$ is given by
\begin{eqnarray}
\mathcal{R}^n[F_{a}]\approx \mathcal{R}^n[\varPsi]+(a-a_{\infty})\mathcal{J}_{n}[\psi],\label{eqA:Claim}
\end{eqnarray}
where $\mathcal{J}_{n}=\mathcal{L}_{\mathcal{R}^{n-1}[\varPsi]}\circ \mathcal{L}_{\mathcal{R}^{n-2}[\varPsi]} \circ \dots \circ \mathcal{L}_{\mathcal{R}^0[\varPsi]}$ and, of course, $\mathcal{R}^0[\varPsi]=\varPsi$.
{To deduce \eqref{eqA:Claim},} let us define $Q_k = \mathcal{R}^k[\varPsi]+(a-a_{\infty})\mathcal{J}_{k}[\psi]$. {For $n=1$ the statement \eqref{eqA:Claim} {reduces to} $\mathcal{R}[F_{a}] \approx Q_1$, which is trivially satisfied in view of \eqref{eqA:RFRF}.} Now, assuming that $\mathcal{R}^k[F_a]\approx Q_k$ holds at step $k$, we just need to prove that $\mathcal{R}^{k+1}[F_a]\approx Q_{k+1}$ also holds. 
Using \eqref{eqA:Jdef}, we get
\begin{eqnarray}
\mathcal{J}_{k+1}[\psi] = \mathcal{L}_{\mathcal{R}^{k}[\varPsi]}\left[\mathcal{J}_{k}[\psi]\right] 
\approx\frac{\mathcal{R}\left[\mathcal{R}^{k}[\varPsi]+(a-a_{\infty})\mathcal{J}_{k}[\psi]\right]-\mathcal{R}\left[\mathcal{R}^{k} [\varPsi]\right]}{a-a_{\infty}},\label{eqA:JKL}
\end{eqnarray}
which implies $\mathcal{R}[Q_{k}] \approx \mathcal{R}^{k+1}[\varPsi]+(a-a_{\infty})\mathcal{J}_{k+1}[\psi] = Q_{k+1}$.
Operating $\mathcal{R}$ on the induction step assumption yields $\mathcal{R}^{k+1}[F_a]\approx \mathcal{R}[Q_k] \approx Q_{k+1}$, which completes the proof of \eqref{eqA:Claim}.

The leading eigenvalue of the eigenvalue problem $\mathcal{L}_g[\varphi]=\delta \varphi$ is $\delta$ (i.e. Feigenbaum's first constant) with associated eigenfunction $\varphi$.
For sufficiently large $n$, the relation
\begin{eqnarray}
\mathcal{J}_{n}[\psi] \approx \delta^n B \varphi \label{eqA:ApproxPhi},
\end{eqnarray}
with $B$ some constant, is expected to hold, because $\mathcal{J}_{n}$ results from the repeated application, many times, of $\mathcal{L}_g$, which amplifies the most unstable mode. 
The eigenvalue problem can be recast in terms of the universal functions as $\mathcal{L}_G[\varPhi]=\delta\varPhi$, where
\begin{eqnarray}\label{PPPP}
\varPhi(\xi)=\mu B\varphi(\gamma \xi)/\gamma
\end{eqnarray}
is normalised to $\varPhi(0)=1$ choosing the scaling factor appropriately as $\mu=\gamma/B\varphi(0)$.


Substituting \eqref{eqA:Gapprox} and \eqref{eqA:ApproxPhi} in \eqref{eqA:Claim} yields
\begin{eqnarray}
\mathcal{R}^n[F_{a}](y)\approx g(y)+(a-a_{\infty})\delta^nB\varphi(y).\label{eqA:Claim2}
\end{eqnarray}
Further replacing $g$ and $\phi$ using \eqref{GGGG} and \eqref{PPPP}, respectively, we obtain
\begin{eqnarray}
\mathcal{R}^n[F_a](y)=\alpha^n F^{2^n}_{a}\left (\frac{y}{\alpha^n} \right)\approx \gamma \left \{G\left (\frac{y}{\gamma} \right) + \delta^n (a-a_{\infty})\mu^{-1}\varPhi \left (\frac{y}{\gamma} \right) \right\},\label{eqA:FG}
\end{eqnarray}
where we have made explicit the effect of the renormalisation operator upon repeated application.
In order to get back to the original map, we can use in \eqref{eqA:FG} the easily verifiable proposition that repeated application of the original and shifted maps is related by $F_{a}^n(y)=f_{a}^n(y+X)-X$ for any $n\in \Natural$. Combined with the definition of the universal function \eqref{eq:UMMM}, the approximation
\begin{eqnarray}
 f^{2^n}_{a}(x) \approx X+\Gamma_n U_M\left (\frac{x-X}{\Gamma_n} \right ),\label{eqA:approx1}
\end{eqnarray} 
follows directly, where the rescaled parameter $M$ and magnification rate of the variable $\Gamma_n$ have been defined as
\begin{eqnarray}\label{eqA:scaled}
M=\delta^n \mu^{-1} (a-a_{\infty})\;\;\text{and}\;\; \Gamma_n=\gamma/\alpha^n.
\end{eqnarray} 

Let us now consider what conditions are necessary for the approximation (\ref{eqA:approx1}) to be valid, besides the one requiring that $2^n$ must be large. The reliability of (\ref{eqA:approx1}) depends on the degree of trust one can place on the approximations \eqref{eqA:Gapprox} and \eqref{eqA:ApproxPhi}. 
It is important to note that, in renormalisation theory, comparisons of functions are conducted on magnified scales. That is, while \eqref{eqA:Gapprox} may be a good approximation on the scale where $y/\alpha^n=(x-X)/\alpha^n$ is not too large, it may not hold true on the original scale where $y\sim O(1)$ (mathematically, this can be easily understood by considering the case where $\varPsi$ is topologically conjugate to $g$). 
In (\ref{eqA:approx1}), {it suffices} to consider $U_M(\xi)$ with the argument in the range $\xi\in[-1,1]$, which encompasses the entire period doubling cascade of \eqref{eq:1dsystemxi}. Hence the restriction $|(x-X)/\Gamma_n| < 1$, which makes \eqref{eqA:Gapprox} only locally valid. The approximation \eqref{eqA:ApproxPhi} is, again, local. Moreover, in order to use the approximation \eqref{eqA:JKL} at level $k=n$, $|\delta^n(a-a_{\infty})|$ must necessarily be small, so we require $|\delta^n M|\ll 1$.



The approximation \eqref{eqA:approx1} must also hold for $U_M(\xi)$ itself, as it is unimodal in the range $\xi\in [-1,1]$. We can therefore set $n=m$, $x=\tilde{x}$, $a=\tilde{M}$, $X=0$, $a_{\infty}=0$, $\mu=\tilde{\mu}$, $\gamma=\tilde{\gamma}$ and $f_a=U_{\tilde{M}}$ in \eqref{eqA:approx1}-\eqref{eqA:scaled} to obtain
\begin{eqnarray}\label{eqA:approx2}
U^{2^m}_{\tilde{M}}(\tilde{x}) \approx  \tilde{\Gamma}_m U_M\left (\frac{\tilde{x}}{\tilde{\Gamma}_m} \right ),\qquad M= \delta^m \tilde{M}/\tilde{\mu},\qquad \tilde{\Gamma}_m=\tilde{\gamma}/\alpha^m,
\end{eqnarray}
which should work as long as $2^m$ is large, $|\tilde{x}/\tilde{\Gamma}_m|<1$, and $|\delta^m \tilde{M}|\ll 1$. 

Finally, particularising \eqref{eqA:approx1} for $n=\ell+m$, we get 
\begin{eqnarray}
\frac{f^{2^{\ell+m}}_{a}(x)-X}{\Gamma_{\ell+m}} \approx  
 U_{M} \left ( \frac{x-X}{\Gamma_{\ell+m}} \right)\approx \frac{\alpha^m}{\tilde{\gamma}}U_{\tilde{M}}^{2^m}\left (\frac{x-X}{\gamma \tilde{\Gamma}_{\ell}} \right),
\end{eqnarray}
where the rightmost approximation is found by setting $\tilde{x}=(\tilde{\gamma}/\gamma)(x-X)\alpha^{-\ell}$ in \eqref{eqA:approx2}. The parameters $a$ and $\tilde{M}$ are linked by $\tilde{M}=\tilde{\mu}\delta^{-m}M=(\tilde{\mu}/\mu)\delta^{\ell}(a-a_{\infty})$. We can simplify the notation by renaming ($\gamma/\tilde{\gamma}$) and ($\mu/\tilde{\mu}$) as $\gamma$ and $\mu$, respectively.
In summary, the approximation
\begin{eqnarray}
\frac{f^{2^{\ell+m}}_{a}(x)-X}{\Gamma_{\ell}} \approx  
 U_{M}^{2^m} \left ( \frac{x-X}{\Gamma_{\ell}} \right),\label{eqA:approx3}
\end{eqnarray}
is obtained if we redefine
\begin{eqnarray}
M = \delta^{\ell}\mu^{-1}(a-a_{\infty}), \qquad \Gamma_\ell=\gamma/\alpha^\ell.
\end{eqnarray}
This approximation is valid for $2^m$ large, irrespective of the value of $\ell=0,1,2\dots$, as long as
\begin{eqnarray}
\left |\frac{x-X}{\Gamma_{\ell+m}} \right | < 1,\qquad |\delta^{\ell+m}\mu^{-1}  (a-a_{\infty})|\ll 1.\label{condition}
\end{eqnarray}
The approximation \eqref{eq:approx} and conditions \eqref{eq:condition} quoted in the main text correspond to the case $\ell=0$.

{
The self-similarity of the period-doubling cascade follows from the validity of the approximation \eqref{eqA:approx3}. Suppose we detect a period-$2^m$ stable orbit in \eqref{eq:1dsystemxi} for $M$ negative but close to zero. Then the relation \eqref{eqA:approx3} asserts the presence of period-$2^{\ell+m}$ orbits in \eqref{eq:1dsystem} at $a=a_\ell=\delta^{-\ell}\mu M+a_{\infty}$ for every $\ell\in \Natural$. Moreover, the orbits must share the same multiplier, because an argument {analogous} to \eqref{eq:diff} can be applied.}

{
The same self-similarity theory applies also to values of the parameter beyond the accumulation point. For instance, suppose we fix $M$ at a small positive number so that a period $2^m p$ window is observed in \eqref{eq:1dsystemxi}, with $p$ an odd number. Then we can expect a period $2^{\ell+m} p$ window in \eqref{eq:1dsystem} at $a_\ell= \delta^{-\ell}\mu M+a_{\infty}$ for every $\ell\in \Natural$. 
Even if the dynamics are predominantly chaotic, we can still expect similarity of the dynamics in the statistical sense. In particular, if $\lambda(M_*)$ is the Lyapunov exponent of \eqref{eq:1dsystemxi} at $M=M_*$, the Lyapunov exponent of \eqref{eq:1dsystem} at $a_\ell=\delta^{-\ell}\mu M_*+a_{\infty}$ should approximately be $2^{-\ell}\lambda(M_*)$. This is consistent with the observation by \citet{HuRu80}. Note that this implies that for all unimodal maps, the distribution of Lyapunov exponents is identical {immediately past} the accumulation point.}

\bibliography{local}
\bibliographystyle{jfm}

\end{document}